\documentclass[aps,prl,reprint,superscriptaddress]{revtex4-1}

\usepackage{graphicx}
\usepackage{physics}
\usepackage{xcolor}

\begin{document}


\title{Anomalous Behavior of Magnetic Susceptibility Obtained by Quench Experiments\\
in Isolated Quantum Systems}


\author{Yuuya Chiba}
\email{chiba@as.c.u-tokyo.ac.jp}
\affiliation{Komaba Institute for Science, The University of Tokyo, 3-8-1 Komaba, Meguro, Tokyo 153-8902, Japan}
\affiliation{Department of Basic Science, The University of Tokyo, 3-8-1 Komaba, Meguro, Tokyo 153-8902, Japan} 

\author{Kenichi Asano}
\email{asano@celas.osaka-u.ac.jp}
\affiliation{Center for Education in Liberal Arts and Sciences, Osaka University, Toyonaka, Osaka 560-0043, Japan}

\author{Akira Shimizu}
\email{shmz@as.c.u-tokyo.ac.jp}
\affiliation{Komaba Institute for Science, The University of Tokyo, 3-8-1 Komaba, Meguro, Tokyo 153-8902, Japan}
\affiliation{Department of Basic Science, The University of Tokyo, 3-8-1 Komaba, Meguro, Tokyo 153-8902, Japan} 

\date{\today}

\begin{abstract}
We examine how the magnetic susceptibility obtained by the quench 
experiment on 
isolated quantum systems is related to the isothermal and adiabatic susceptibilities defined in thermodynamics.
Under the conditions similar to the eigenstate thermalization hypothesis, together with some additional natural ones, 
we prove that for translationally invariant systems the quench susceptibility as a function of wave vector $\vb*{k}$ is discontinuous at $\vb*{k}=\vb*{0}$. 
Moreover, its values at $\vb*k=\vb*0$ and the $\vb*k\to\vb*0$ limit coincide with the adiabatic and the isothermal susceptibilities, respectively.
We give numerical predictions on how these particular behaviors can be observed in experiments on the $XYZ$ spin chain with tunable parameters, and how they deviate when the conditions are not fully satisfied. 
\end{abstract}


\maketitle



\paragraph{\label{par:intro}Introduction.---}%
Ultracold atoms \cite{Lewenstein2007,Bloch2008} and molecules \cite{Micheli2006,Chin2009,Carr2009} in optical lattices offer nearly ideal playgrounds for studying quantum many-body systems experimentally. 
Various model systems 
\cite{Jaksch1998,Greiner2002a,Stoferle2004,Spielman2007,Spielman2008,Kohl2005,Parsons2016,Esslinger2010,Wall2015,Pelegri2019,Simon2011,Fukuhara2013b,Fukuhara2015,Yan2013,Hazzard2014,Orioli2018} 
are realized on the optical lattices with various geometry 
\cite{Amico2005,Henderson2009,Wirth2011,Soltan-Panahi2011,Tarruell2012,Jo2012} 
and with tunable physical parameters \cite{Feshbach1958,Fano1961,Tiesinga1993,Stoof1996,Bloch2008}. 
Furthermore, 
one can isolate the systems from the environments over a reasonably long period, 
which enables the direct observation of the dynamics of isolated quantum systems 
induced by suddenly changing a physical parameter 
\cite{Greiner2002b,Sadler2006,Meinert2013,Lamporesi2013,Hung2013,Fukuhara2013a, Hild2014}.
After this so-called quench, the system often relaxes to a steady state, 
where the expectation values of local observables become almost time independent 
\cite{Trotzky2012, Orioli2018,  Kinoshita2006,Gring2012,Tasaki1998,Reimann2008,Mori2018}. 
The nature of such a steady state 
has been discussed in terms of 
the eigenstate thermalization hypothesis (ETH)
\cite{VonNeumann2010,Deutsch1991,Srednicki1994,Rigol2008,Srednicki1999,DAlessio2016,Anza2018,Goldstein2017,Biroli2010,Iyoda2017,Mori2016,Kuwahara2019}.
For example, if the ``strong'' ETH 
is satisfied, 
the steady state is an equilibrium state \cite{VonNeumann2010,Deutsch1991,Srednicki1994,Rigol2008,Srednicki1999,DAlessio2016,Anza2018}.

In this Letter, we 
study the susceptibility obtained by the {\em quench} experiment, 
and explore whether or not it coincides with a thermodynamic susceptibility. 
This problem is highly nontrivial 
since there are two kinds of thermodynamic susceptibilities, 
the {\em isothermal} and the {\em adiabatic} ones,
which take different values. 
In other words, it is not even clear which thermodynamic susceptibilities should be compared with the quench one. 
Furthermore, 
the wave number dependences of 
these susceptibilities make the problem more nontrivial, as we will reveal in this paper.

To be concrete, we consider the magnetic susceptibility of a quantum spin system.
Suppose that the initial equilibrium state is in a uniform ``offset'' magnetic field $h$,
and a weak extra magnetic field of wave number $\vb*{k}$ is suddenly applied.
The {\em quench susceptibility} $\chi^{\rm qch}(\vb*{k})$ is defined as the rate of magnetization change 
induced by such a quench.
We explore its relation to 
the isothermal and the adiabatic thermodynamic susceptibilities, 
$\chi^{T}(\vb*{k})$ and $\chi^{S}(\vb*{k})$,
in the case where $\chi^{T}(\vb*{0})>\chi^{S}(\vb*{0})$,
which occurs when $h \neq 0$.

We reveal that 
$\chi^{\rm qch}(\vb*{k})$ 
is {\em discontinuous} at $\vb*{k}=\vb*{0}$ as a function of $\vb*{k}$. 
Because of this discontinuity, 
{\em both} thermodynamic susceptibilities are obtained from the quench one,
as $\chi^{\text{qch}}(\vb*{0})=\chi^{S}(\vb*0)$ and $\displaystyle{\lim_{\vb*{k}\to\vb*{0}}}\chi^{\text{qch}}(\vb*{k})=\chi^{T}(\vb*0)$. 
The proof requires the conditions similar to the ETH, 
which hold when the dynamics of the system is complicated enough, 
as well as the natural conditions that are satisfied except at a phase transition point. 

Furthermore, we numerically demonstrate how such anomalous behaviors 
should be observed in experiments on an isolated quantum spin system 
when it is nonintegrable.
We also predict how the deviation from these behaviors is observed when the physical parameters of the system are tuned so that it becomes integrable.

\paragraph{Setup.---}
We deal with a quantum spin-$1/2$ system 
on a $d$-dimensional cubic lattice $\Omega_{N}$ with linear size $L$ 
and $N=L^d$ spins.
The periodic boundary conditions and the invariance under the discrete spatial translations are assumed for the prequench Hamiltonian $\hat{H}(h)$, 
where $h$ denotes the uniform offset magnetic field. 
The density matrix of the initial state is chosen as the canonical Gibbs one, 
$\hat{\rho}_{\rm ini}=e^{-\beta \hat{H}(h)}/Z$
\footnote{The initial state can be replaced with an appropriate pure quantum state that approximates the state in Ref.~\cite{Sugiura2013}, which gives the same results as the Gibbs state for both equilibrium \cite{Sugiura2013} and dynamical \cite{Shimizu2017,Endo2018} properties.}. 

We are interested in the {\em quantum quench process} 
where the additional magnetic field $\Delta h(\vb*{r})$,
with wave number $\vb*{k}$ and small magnitude $\Delta h_{\vb*{k}}$,
is suddenly applied  at $t=0$. 
At $t > 0$, the isolated system obeys the Schr\"{o}dinger dynamics of the postquench Hamiltonian, 
$\hat{H}(h)-\sum_{\vb*{r}\in\Omega_N}\hat{\sigma}_{\vb*{r}}^z\Delta h(\vb*{r})$,
where $\hat{\sigma}^{\alpha}_{\vb*{r}}$ ($\alpha = x,y,z$) is the Pauli operator on site $\vb*{r}\in\Omega_{N}$.
For simplicity, we assume that $\Delta h(\vb*{r})$ is parallel to the offset field $h$, pointing in the $z$ direction.
While the previous works regarding the quantum quench focused only on the final state, 
we here study the {\em quench susceptibility}, 
\begin{align}
\chi_N^{\text{qch}}(\vb*{k}):=\lim_{\mathcal{T}\to \infty}\lim_{\Delta h_{\vb*{k}}\to 0}
\frac{\overline{\text{Tr}[ \hat{\rho}(t)\hat{m}_{\vb*{k}}]}^{\mathcal{T}}-\text{Tr}[ \hat{\rho}_{\rm ini}\hat{m}_{\vb*{k}}]}{\Delta h_{\vb*{k}}}, 
\label{eq:chi^qch}
\end{align}
which quantifies the {\em difference} of the expectation values of the $\vb*k$ component of magnetization, 
$\hat{m}_{\vb*{k}}=(1/N)\sum_{\vb*{r}\in\Omega_{N}} e^{-i\vb*{k}\vdot\vb*{r}}\hat{\sigma}^z_{\vb*{r}}$, 
between the final and the initial states. 
Here, $\hat{\rho}(t)$ is the density matrix at time $t$, and $\overline{f(t)}^{\mathcal{T}}$ denotes the time average of $f(t)$ over $0\le t \le \mathcal{T}$.

For comparison, 
we consider the isothermal and the adiabatic thermodynamic susceptibilities, 
$\chi^{T}_N(\vb*{k})$ and $\chi^{S}_N(\vb*{k})$, 
which are defined via
the quasistatic processes
with constant temperature and entropy, 
respectively.
At $\vb*{k}=\vb*{0}$, they satisfy 
\begin{align}
\chi_N^S(\vb*{0})
=
\chi_N^T(\vb*{0})
-
\frac{T}{c_h}\left[ \left( \frac{\partial m_{\vb*{0}}}{\partial T} \right)_h \right]^2,
%
\label{eq:chi_relation}
\end{align}
where $c_h$ is the specific heat at constant magnetic field and $T=1/\beta$ is the temperature \footnote{See Supplemental Material for detailed calculations, which includes Refs.~\cite{Mazur1969, Beugeling2014, Steinigeweg2014}}.
We assume $0<T< +\infty$,
and exclude phase transition points where $c_h$ diverges 
as $N \to \infty$ 
and the case where 
$({\partial m_{\vb*{0}}}/{\partial T})_{h}$ vanishes, 
which is indeed unlikely for $h \neq 0$.
This leads to the most interesting situation where 
the two susceptibilities take different values
even in the thermodynamic limit,
\begin{align}
\chi^{S}_{\infty}(\vb*{0}) < \chi^{T}_{\infty}(\vb*{0}), 
\label{chiT>chiS}
\end{align}
where $\chi_{\infty}^{\bullet}(\vb*{k})
:=
\lim_{N\to\infty}\chi_{N}^{\bullet}(\vb*{k})$.

\paragraph{Main results.---}
Our main results are summarized as follows.

(i) The $\vb*{k}=\vb*{0}$ value of the quench susceptibility agrees
with that of the adiabatic one:
\begin{align}
\chi_{\infty}^{\text{qch}}(\vb*{0})=\chi_{\infty}^{S}(\vb*{0}),\label{eq:qch=S}
\end{align}
if and only if condition (\ref{eq:weakETHlike}), 
which is similar to but different from the ordinary ETH, 
is satisfied. 
Although the quench increases entropy, 
this equality implies it is irrelevant to $\chi_{\infty}^{\text{qch}}(\vb*{0})$ \cite{Note2}.
By contrast, 
the quench induces relevant changes in energy and temperature \cite{Note2},
which results in $\chi_{\infty}^{\text{qch}}(\vb*{0})<\chi_{\infty}^{T}(\vb*{0})$.

(ii) The $\vb*{k}\neq \vb*{0}$  value of the quench susceptibility agrees with those of the adiabatic and the isothermal ones \footnote{The equality, $\chi_{N}^S(\vb*{k})=\chi_{N}^T(\vb*{k})$ for all $\vb*{k}\neq\vb*{0}$ and for all $N$, can be proved only from the translation invariance \cite{Note2}.},
\begin{align}
\chi_{\infty}^{\text{qch}}(\vb*{k})=\chi_{\infty}^{S}(\vb*{k})=\chi_{\infty}^{T}(\vb*{k})\text{  for all }\vb*{k}\neq \vb*{0}\label{eq:knon0},
\end{align}
if and only if condition (\ref{eq:offdETHlike}), which is similar to but weaker than the ordinary ``off-diagonal'' ETH \cite{Srednicki1999,DAlessio2016,Mori2018,Anza2018}, is satisfied.

(iii) The isothermal susceptibility, $\chi_{\infty}^{T}(\vb*{k})$, 
is {\em uniformly} continuous as a function of $\vb*k$ \footnote{See Refs.~\cite{Lebowitz1967,Hansen2013} for related discussions.}
under two conditions (\ref{eq:phi_decay}) and (\ref{eq:phi_conv}) regarding the spatial spin-spin correlation function, both of which are fulfilled in normal systems.

(iv) When the conditions for 
(ii) and (iii), 
[namely, Eqs.~(\ref{eq:offdETHlike}), (\ref{eq:phi_decay}) and (\ref{eq:phi_conv})]
 are all satisfied, 
\begin{align}
\lim_{\vb*{k}\to\vb*{0}}\chi_{\infty}^{\text{qch}}(\vb*{k})
= \lim_{\vb*{k}\to\vb*{0}}\chi_{\infty}^{T}(\vb*{k})
=\chi_{\infty}^{T}(\vb*{0}).
\label{eq:qch=T}
\end{align}
This also shows 
that $\chi_{\infty}^{\text{qch}}(\vb*{k})$ is discontinuous at $\vb*k=\vb*0$ 
because
$\chi_{\infty}^{\text{qch}}(\vb*0) < \chi_{\infty}^T(\vb*0)$ as seen 
from the thermodynamic inequality (\ref{chiT>chiS}) 
\footnote{If $h=0$, so that $m_{\vb*{0}}=0$, we have 
$\chi_{\infty}^{S}(\vb*{0})=\chi_{\infty}^{T}(\vb*{0})$
unlike (\ref{chiT>chiS}).
Even in such a case, Eq.~(\ref{eq:qch=T}) shows that $\chi_{\infty}^{\text{qch}}(\vb*{k})$ is discontinuous at $\vb*k=\vb*0$ unless the condition for (i) is satisfied.}
and the general relation \cite{Note2},
\begin{equation}
\chi_N^{\text{qch}}(\vb*{0})\le\chi_N^{S}(\vb*{0}).
\label{inequality_chi}
\end{equation}

(v) These results can be confirmed by a series of experiments in the isolated quantum systems, e.g., ultracold atoms, which simulate the $XYZ$ spin chain. 
We predict the dependence of the above susceptibilities on $\vb*{k}$, $N$, and the exchange coupling parameters, $J_x, J_y, J_z$.

\paragraph{Condition for (i).---}
We introduce
$\hat{m}_{\vb*{k}}^0:=\lim_{\mathcal{T}\to\infty}\overline{e^{i\hat{H}(h)t}\hat{m}_{\vb*{k}}e^{-i\hat{H}(h)t}}^{\mathcal{T}}$, which is the energy-diagonal part of $\hat{m}_{\vb*{k}}$ \cite{Note2}.
Let $\ket{\nu}$ be the simultaneous eigenstate 
of $\hat{H}(h)$, the translation operators, and $\hat{m}_{\vb*{k}=\vb*{0}}^0$, 
with eigenenergy $E_{\nu}$ and crystal momentum $\vb*{K}_{\nu}$. 
We also introduce 
$\delta\hat{\sigma}^z_{\vb*{r}}=\hat{\sigma}^z_{\vb*{r}}-\text{Tr}[\hat{\rho}_{\rm ini}\hat{\sigma}^z_{\vb*{r}}]$ 
and $\delta E_{\nu}= E_{\nu}-\text{Tr}[\hat{\rho}_{\rm ini}\hat{H}(h)]$.
Then, we obtain the necessary and sufficient condition for (i) in the following form \cite{Note2}:
For {\em almost all} $\ket{\nu}$ 
in a {\em narrow} energy region $|\delta E_{\nu}| \lesssim T\sqrt{c_h N}$, 
the diagonal elements $\mel{\nu}{\delta \hat{\sigma}^z_{\vb*{0}}}{\nu}$ are 
related almost linearly with $\delta E_{\nu}$ as 
\begin{equation}
\mel{\nu}{\delta \hat{\sigma}^z_{\vb*{0}}}{\nu}= C \ \delta E_{\nu}/N \ +o(1/\sqrt{N}), 
\label{eq:weakETHlike}
\end{equation}
where $C=\mathcal{O}(1)$ is some constant independent of $\nu$ 
\footnote{Note that condition~(\ref{eq:weakETHlike}) can be satisfied even when $C=0$, although $C$ do not vanish in our setting where $h\neq 0$ \cite{Note2}.}.
This is similar to but different from the ordinary two forms of ETH in the following points. 
The ordinary strong ETH \cite{Srednicki1999,Rigol2008,DAlessio2016,Anza2018} requires more stringently 
that {\it all} $\mel{\nu}{\hat{\sigma}^z_{\vb*{0}}}{\nu}$ 
behave like a smooth function of $E_{\nu}/N$, 
which is often satisfied in nonintegrable 
systems \cite{Kim2014}.
Since a smooth function of $E_{\nu}/N$ can be regarded as linear 
within the narrow region $|\delta E_{\nu}|\lesssim T\sqrt{c_h N}$,
any system satisfying the strong ETH also satisfies condition (\ref{eq:weakETHlike}).
By contrast, the ordinary weak ETH \cite{Biroli2010,Iyoda2017,Mori2016} requires only that $\mel{\nu}{\delta \hat{\sigma}^z_{\vb*{0}}}{\nu}=o(1)$ for almost all $\nu$ in the same energy region. 
For this reason, some models that satisfy the ordinary weak ETH do not satisfy Eq.~(\ref{eq:weakETHlike}), as will be demonstrated shortly. 


\paragraph{Demonstration of (i).---}
We now demonstrate how result (i) can be observed in experiments on the $XYZ$ spin chain, which has the prequench Hamiltonian, 
\begin{align}
\hat{H}(h)=-\sum_{j=0}^{N-1} \sum_{\alpha=x,y,z}J_{\alpha}\hat{\sigma}_{j}^{\alpha}\hat{\sigma}_{j+1}^{\alpha}-\sum_{j=0}^{N-1} h\hat{\sigma}_{j}^z,
\label{eq:H}
\end{align}
with periodic boundary condition, $\hat{\vb*\sigma}_N=\hat{\vb*\sigma}_0$. 
Since spin systems \cite{Wall2015,Pelegri2019,Simon2011,Fukuhara2013b,Fukuhara2015,Yan2013,Hazzard2014,Orioli2018} and a 1D ring \cite{Amico2005,Henderson2009} can be separately realized in ultracold atoms and molecules, 
we expect this model can also be realized experimentally. 
This model alone covers three different classes of systems, 
(a) $XYZ$, (b) $XXZ$ ($J_x=J_y \neq J_z$), and (c) $XY$ ($J_z=0$) models, 
by tuning the parameters $J_\alpha$. 
We here predict the behaviors of the susceptibilities by means of the numerical diagonalization for (a) and (b), and the analytic evaluation for (c), respectively.

\begin{figure}[]
\includegraphics[width=8.6cm]{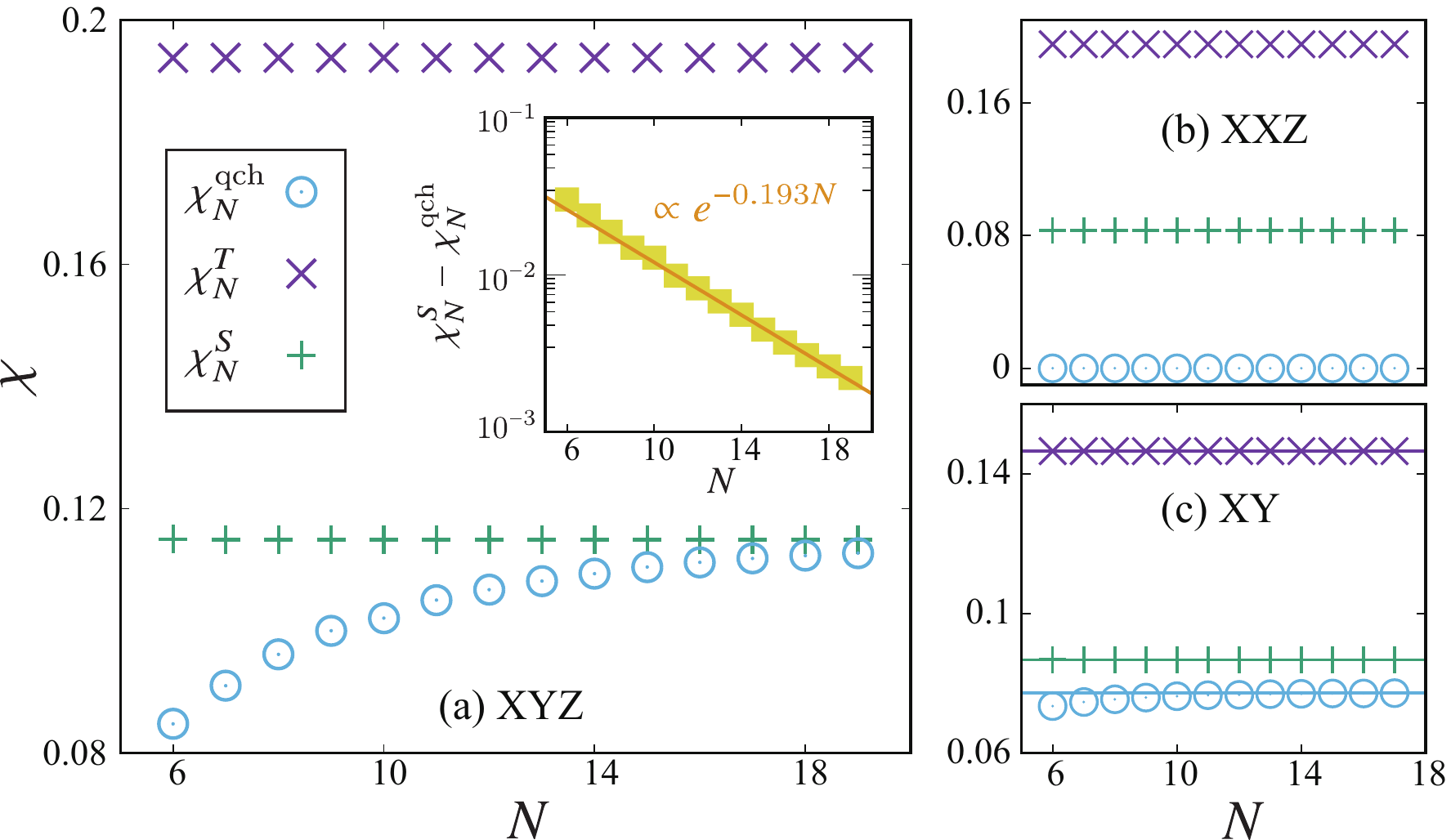}%
\caption{\label{fig:N}
Size-$N$ dependence of $\chi^{\text{qch}}_N(0)$, $\chi^{T}_N(0)$, and $\chi^{S}_N(0)$ of 
the (a) $XYZ$, (b) $XXZ$, and (c) $XY$ models. 
We take (a) $(J_x-J_y, J_z)=(1.2, 1.0)$,
(b) $(0.0, 1.0)$, and (c) $(1.2, 0.0)$, 
for fixed values of $J_x+J_y =0.6, h=0.8, \beta=0.15$. 
Inset of (a) : $\chi^{S}_N(0) - \chi^{\text{qch}}_N(0)$ in the logarithmic scale.
Solid lines in (c): $\chi^{\text{qch}}_{\infty}(0)$, $\chi^{T}_{\infty}(0)$, and $\chi^{S}_{\infty}(0)$.
}
\end{figure}

Figure \ref{fig:N}(a) shows the $N$ dependence of the 
$k=0$ components $\chi^{\text{qch}}_N(0)$, $\chi^{T}_N(0)$, and $\chi^{S}_N(0)$ in the $XYZ$ model 
\footnote{
Here, we take $J_y$ negative in (a) because we found that 
the $o(1/\sqrt{N})$ term of Eq.~(\ref{eq:weakETHlike}) is smaller for larger $J_x-J_y$.}.
Since the model has no local conserved quantity for $h \neq 0$ 
\cite{Shiraishi2019}, 
it is expected that the condition~(\ref{eq:weakETHlike}) is fulfilled, so that Eq.~(\ref{eq:qch=S}) holds.
In fact, Fig.~\ref{fig:N}(a) shows that 
$\chi^{\text{qch}}_N(0)$ approaches $\chi^{S}_N(0)$ as $N$ increases. 
Their difference decreases nearly exponentially, as shown in the inset,
where the function $0.083 \ e^{-0.193N}$ is also plotted as a guide to the eye.
Both of them remain far off from $\chi^{T}_N(0)$. 

Contrastingly, Eq.~(\ref{eq:qch=S}) does {\em not} hold for the $XXZ$ or the $XY$ models, 
as shown in Figs.~\ref{fig:N}(b) and \ref{fig:N}(c), respectively. 
In these two cases, there exist some local conserved quantities 
that result in the violation of Eq.~(\ref{eq:weakETHlike}) and its equivalent (\ref{eq:qch=S}). 
In other words, they do not satisfy Eq.~(\ref{eq:weakETHlike}) because of its integrability \cite{Alba2015}, 
while they {\it do satisfy} the ordinary ``weak'' ETH \cite{Biroli2010,Iyoda2017,Mori2016,Kuwahara2019}. 
It should be noted that our results (a)-(c) are consistent with 
inequalities (\ref{chiT>chiS}) and (\ref{inequality_chi}).

\paragraph{Conditions for (ii).---}
As is proved in Ref.~\cite{Note2}, Eq.~(\ref{eq:knon0}) holds 
if and only if 
{\em almost all} $\ket{\nu}$ in a {\em narrow} 
energy region $|\delta E_{\nu}|\lesssim T\sqrt{c_h N}$ satisfy
\begin{align}
\sum_{\nu'}
&\delta_{E_{\nu},E_{\nu'}}\delta_{\vb*{K}_{\nu},\vb*{K}_{\nu'}+\vb*{k}}
|\mel{\nu'}{\hat{\sigma}^z_{\vb*{0}}}{\nu}|^2 = o(1/N)\notag\\
&\text{  for all }\vb*{k}\neq \vb*{0}.
\label{eq:offdETHlike}
\end{align}
This is similar to the ``off-diagonal ETH'' \cite{Srednicki1999,Goldstein2017,DAlessio2016,Mori2018,Anza2018}, except for the following points. 
First, the off-diagonal ETH requires that {\em all} off-diagonal elements of {\em all} local operators tend to vanish as $N \to \infty$.
By contrast, 
Eq.~(\ref{eq:offdETHlike}) 
refers only to a {\em particular} spin operator $\hat{\sigma}^z_{\vb*{0}}$
and to the off-diagonal elements
between {\em specific} pairs of states such that
\begin{equation}
E_{\nu}=E_{\nu'} \mbox{ and } \vb*{K}_{\nu}=\vb*{K}_{\nu'}+\vb*{k}.
\label{cond:EandK}
\end{equation}
Furthermore, it 
requires not all such off-diagonal elements but {\em most} of them tend to vanish.
Second, 
the ordinary off-diagonal ETH \cite{Srednicki1999,DAlessio2016,Mori2018,Anza2018} requires exponentially fast decay of all the off-diagonal elements, which is not necessarily satisfied in integrable models. 
By contrast, Eq.~(\ref{eq:offdETHlike}) is a weaker condition \cite{Note2} that can be satisfied even 
in integrable models, as we will demonstrate shortly for the $XY$ model. 

It is noteworthy that if we impose Eqs.~(\ref{eq:weakETHlike}) and (\ref{eq:offdETHlike}) not only on a {\em particular} spin operator $\hat{\sigma}^z_{\vb*{0}}$ but also on all other local operators, we obtain a 
{\em new necessary condition for thermalization},
which is also a sufficient condition as long as the quench parameter 
$\Delta h_{\vb*{k}}$ is small. 

\paragraph{Conditions for (iii).---} 
We introduce the canonical spin-spin correlation function \cite{KTH,Note2} as $\phi_N^{T}(\vb*{r}):=\beta\langle \delta\hat{\sigma}^z_{\vb*{0}};\delta\hat{\sigma}^z_{\vb*{r}} \rangle_{\rm ini}$. 
Then, we can show \cite{Note2} that $\chi^{T}_{\infty}(\vb*{k})$ is {\em uniformly continuous on the whole region} ({\em including} $\vb*{k}=\vb*{0}$), if $\phi^{T}_{\infty}(\vb*{r})$ decays fast enough such that
\begin{align}
\lim_{N\to\infty} \sum_{\vb*{r}\in\Omega_{N}}
\left| \phi^{T}_{\infty}(\vb*{r}) \right|
< \infty
\label{eq:phi_decay}
\end{align}
and if finite-size effects are small such that 
\begin{align}
\lim_{N\to \infty} \sum_{\vb*{r}\in\Omega_{N}}
\left| \phi_N^{T}(\vb*{r})-\phi^{T}_{\infty}(\vb*{r}) \right|
=0.
\label{eq:phi_conv}
\end{align}
Since we exclude phase transition points, 
condition (\ref{eq:phi_decay}) is expected to be satisfied in most systems.
Moreover, 
it seems normal that the condition (\ref{eq:phi_conv}) holds, since the canonical ensemble
well emulates a subsystem in an infinite system \cite{Hyuga2014,Iyer2015}.

If conditions (\ref{eq:offdETHlike}), (\ref{eq:phi_decay}) and (\ref{eq:phi_conv}) 
are all fulfilled,
Eq.~(\ref{eq:qch=T})
follows from results (ii) and (iii).
It also follows that 
$\chi_{\infty}^{\text{qch}}(\vb*{k})$ is discontinuous at $\vb*k=\vb*0$, 
as discussed in (iv).

\paragraph{Demonstrations of (ii)-(iv).---}
The discontinuity of $\chi^{\text{qch}}_\infty(\vb*{k})$ may seem counterintuitive, 
but can be verified experimentally by adopting the isolated system representing Eq.~(\ref{eq:H}). 
The observed susceptibility 
should follow the following results of the numerical simulation. 

\begin{figure}
\includegraphics[width=8.6cm]{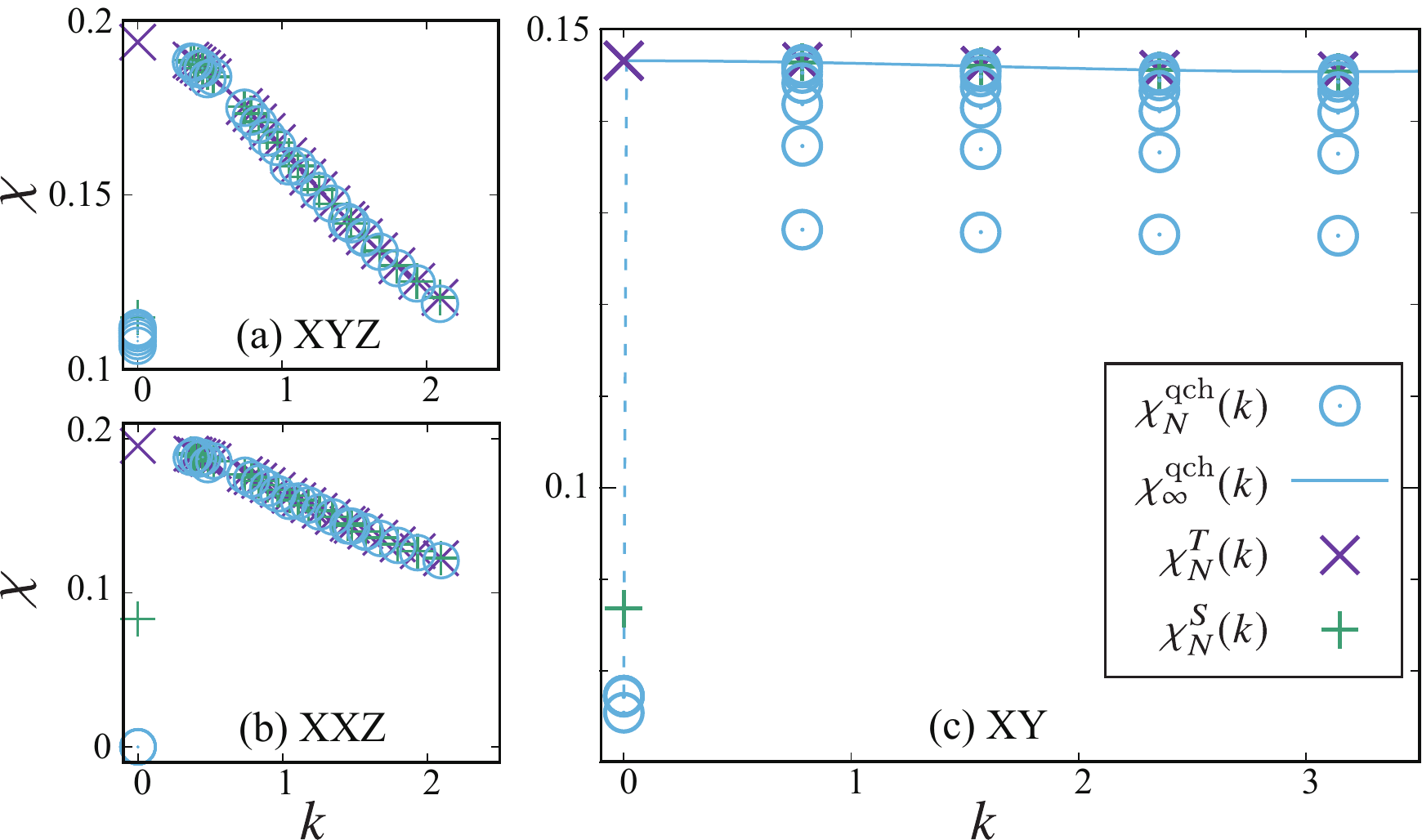}%
\caption{\label{fig:k}$k$ dependence of $\chi^{\text{qch}}_N(k)$, $\chi^{T}_N(k)$, and $\chi^{S}_N(k)$ in (a) $XYZ$, (b) $XXZ$, and (c) $XY$ 
models, with the same parameters as in Fig.~\ref{fig:N}.
We take (a), (b) $N=12$-$17$ and $k=2\pi n_k/N$, 
and (c) $N=2^n$ with $n=3$-$9$ and $k=2\pi n_k/8$,  with $n_k=0$-$4$. 
Solid line in (c): $\chi^{\text{qch}}_{\infty}(k)$ [$=\chi^{S}_{\infty}(k)=\chi^{T}_{\infty}(k)$] for $k\neq 0$, whereas the dashed line shows its discontinuous jump to $\chi^{\text{qch}}_{\infty}(0)$.}
\end{figure}

Figure \ref{fig:k} shows the $k$ dependence of 
$\chi^{\text{qch}}_N(k)$, $\chi^{T}_N(k)$, and $\chi^{S}_N(k)$ 
in the (a) $XYZ$, (b) $XXZ$, and (c) $XY$ models.
Recalling that the condition~(\ref{eq:offdETHlike}) is weaker than the ordinary off-diagonal ETH \cite{Srednicki1999,DAlessio2016,Mori2018,Anza2018}, we expect that it is fulfilled in all these models.
In fact, our data show that Eq.~(\ref{eq:knon0}), 
$\chi_{\infty}^{\text{qch}}(k)=\chi_{\infty}^{S}(k)=\chi_{\infty}^{T}(k)$ 
for all $k \neq 0$, 
holds in each model. 
We also find that $\chi^{T}_N(k)-\chi^{\text{qch}}_N(k)$ for $k\neq 0$ scales as $\Theta(1/N)$ in (c).
This is because the off-diagonal elements $|\mel{\nu'}{\hat{\sigma}^z_{\vb*{0}}}{\nu}|$
that satisfy Eq.~(\ref{cond:EandK}) decay not exponentially but algebraically as $\Theta(1/N)$ for the $XY$ model.

The conditions (\ref{eq:phi_decay}) and (\ref{eq:phi_conv}) are the natural 
ones that will 
also be satisfied in all these models. 
In fact, Figs.~\ref{fig:k}(a)-\ref{fig:k}(c) indicate Eq.~(\ref{eq:qch=T}), $\displaystyle{\lim_{k \to 0}}\chi_{\infty}^{\text{qch}}(k)=\chi_{\infty}^{T}(0)$, holds and hence $\chi_{\infty}^{\text{qch}}(k)$ is discontinuous at $k=0$ while $\chi_{\infty}^{T}(k)$ is uniformly continuous.

For the parameters presented here, Eqs.~(\ref{eq:knon0}) and (\ref{eq:qch=T}) hold in all three cases, 
while Eq.~(\ref{eq:qch=S}) only in the $XYZ$ one. 
By further varying $J_x$ and $J_y$, we can also construct a model for which {\em none} of Eqs.~(\ref{eq:qch=S})-(\ref{eq:qch=T}) holds \cite{Note2}.
In such a case, the condition (\ref{eq:offdETHlike}) is violated, 
while the conditions (\ref{eq:phi_decay}) and (\ref{eq:phi_conv}) are still fulfilled.

\paragraph{Discussion on discontinuity .---}
The discontinuity of 
$\chi_{\infty}^{\text{qch}}(\vb*{k})$ 
at $\vb*{k}=\vb*{0}$
seems nontrivial.
When results (i) and (ii) hold, 
this discontinuity is related to 
that of $\chi_{\infty}^{S}(\vb*{k})$.
To help understand the former,
we here explain the latter discontinuity intuitively \cite{Note4}.
We also explain the continuity of $\chi_{\infty}^{T}(\vb*{k})$.

Suppose that a huge system is enclosed by an adiabatic wall.
We consider 
a thermodynamic process 
in which $\Delta h(\vb*{r})$ is applied quasistatically.
If $\Delta h(\vb*{r})$ is localized and uniform in a subsystem with $N=L^d$ sites,
one obtains the adiabatic susceptibility 
of wave number $|\vb*{k}| \sim 1/L$.
Since the total system size is huge, it is well approximated by $\chi_{\infty}^{S}(\vb*{k})$. 
On the other hand, 
this thermodynamic process can also be regarded as an isothermal process for the subsystem 
because the rest of the system works as a heat reservoir.
According to this picture, 
one obtains $\chi_{N}^{T}(\vb*{0})$.
Since the two pictures have to give the same results, 
$\chi_{\infty}^{S}(\vb*{k})=\chi_{N}^{T}(\vb*{0})$ for $|\vb*{k}| \sim 1/L$.
By increasing $N$, 
we obtain $\lim_{\vb*{k}\to\vb*{0}}\chi_{\infty}^{S}(\vb*{k})=\chi_{\infty}^{T}(\vb*{0})$. 
Comparing this with inequality (\ref{chiT>chiS}), 
we can see that $\chi_{\infty}^{S}(\vb*{k})$ is discontinuous at $\vb*{k}=\vb*{0}$.
By contrast, we can argue similarly the case where the adiabatic wall is replaced with 
a heat reservoir.
Then we have
$\lim_{\vb*{k}\to\vb*{0}}\chi_{\infty}^{T}(\vb*{k})=\chi_{\infty}^{T}(\vb*{0})$,
which shows that $\chi_{\infty}^{T}(\vb*{k})$ is continuous at $\vb*{k}=\vb*{0}$.

\paragraph{Relation to Kubo formula .---} 
Since the Schr\"odinger dynamics is assumed, 
our results are applicable to experiments on isolated quantum systems. 
Moreover, since many formulas of physics were derived assuming 
the Schr\"odinger dynamics, 
our results contribute also to foundations of such formulas. 
As an example of the latter,
we finally discuss 
the susceptibility obtained by the Kubo formula  \cite{Kubo1957}, 
$\chi_N^{\text{Kubo}}(\vb*{k},\omega+i\varepsilon)$.
Here, $\omega$ is the frequency and $\varepsilon$ is an infinitesimal positive number.
While we have defined $\chi_N^{\text{qch}}$ through a sudden quench of 
$\Delta h(\vb*{r})$,
Kubo derived $\chi_N^{\text{Kubo}}$ assuming that $\Delta h(\vb*{r})$ 
is switched on gradually over a long timescale $\sim 1/\varepsilon$.

It is generally believed that the $\varepsilon\to+0$ limit of $\chi_N^{\text{Kubo}}$ 
should be taken {\em after} the $N\to\infty$ limit \cite{Pines1966,Giuliani2005,Zubarev1974,Zubarev1996,Zubarev1997}.
However,  some works took the $\varepsilon\to+0$ limit keeping $N$ finite \cite{Falk1968,Wilcox1968,Suzuki1971}.
For the latter limit, 
we can show \cite{Note2}
\begin{align}
\lim_{\varepsilon\to+0}\chi_N^{\text{Kubo}}(\vb*{k},0+i\varepsilon)=\chi_N^{\text{qch}}(\vb*{k})\quad \mbox{for all $N$},\label{eq:kubo=qch}
\end{align}
although the left-hand side and the right-hand side 
correspond to the slow and fast 
processes, respectively, which would result in different final states. 
Therefore, all the statements (i)-(iv) for $\chi_{\infty}^{\text{qch}}(\vb*{k})$ 
hold also for $\displaystyle \lim_{N\to\infty}\lim_{\varepsilon\to+0}\chi_N^{\text{Kubo}}(\vb*{k},0+i\varepsilon)$ \footnote{
For the former order of limits ($\varepsilon\to+0$ after $N\to\infty$), 
the conventional wisdom \cite{Pines1966,Giuliani2005} is that $\displaystyle\lim_{\vb*{k} \to \vb*{0}} \lim_{\varepsilon \to +0} \lim_{N \to \infty}\chi_N^{\text{Kubo}}(\vb*{k}, 0+i\varepsilon) = \chi_\infty^{T}(\vb*{0})$, which however does not always hold.
Our results (ii)-(iv) suggest the condition for the validity of this wisdom, although we have not yet proved $\displaystyle\lim_{\varepsilon \to +0} \lim_{N \to \infty}\chi_N^{\text{Kubo}}(\vb*{k}, 0+i\varepsilon) = \chi_\infty^{\text{qch}}(\vb*{k})$.
}.
Moreover, 
the previous results on $\displaystyle\lim_{\varepsilon\to+0}\chi_N^{\text{Kubo}}(\vb*{0}, 0+i \varepsilon)$ \cite{Falk1968,Wilcox1968,Suzuki1971} can be understood more precisely using (i) \cite{Note2}.
However, it is noteworthy that $\chi_N^{\text{Kubo}}$ is hard to measure in experiments in contrast 
to $\chi_N^{\text{qch}}$, since the system cannot be isolated for the infinitely long timescale. 

In conclusion, we have revealed the anomalous natures of the quench susceptibility, 
demonstrating together that experimental verifications are feasible enough.

\begin{acknowledgments}
We thank Y.~Yoneta, A.~Noguchi and Y.~Kato for discussions, and 
A. Ikeda, C.~Hotta, R.~Hamazaki and K.~Saito for helpful comments. 
This work was supported by JSPS KAKENHI Grants No.~JP19H01810, No.~JP15H05700, and No.~JP17K05497.
\end{acknowledgments} 


\begin{thebibliography}{87}%
\makeatletter
\providecommand \@ifxundefined [1]{%
 \@ifx{#1\undefined}
}%
\providecommand \@ifnum [1]{%
 \ifnum #1\expandafter \@firstoftwo
 \else \expandafter \@secondoftwo
 \fi
}%
\providecommand \@ifx [1]{%
 \ifx #1\expandafter \@firstoftwo
 \else \expandafter \@secondoftwo
 \fi
}%
\providecommand \natexlab [1]{#1}%
\providecommand \enquote  [1]{``#1''}%
\providecommand \bibnamefont  [1]{#1}%
\providecommand \bibfnamefont [1]{#1}%
\providecommand \citenamefont [1]{#1}%
\providecommand \href@noop [0]{\@secondoftwo}%
\providecommand \href [0]{\begingroup \@sanitize@url \@href}%
\providecommand \@href[1]{\@@startlink{#1}\@@href}%
\providecommand \@@href[1]{\endgroup#1\@@endlink}%
\providecommand \@sanitize@url [0]{\catcode `\\12\catcode `\$12\catcode
  `\&12\catcode `\#12\catcode `\^12\catcode `\_12\catcode `\%12\relax}%
\providecommand \@@startlink[1]{}%
\providecommand \@@endlink[0]{}%
\providecommand \url  [0]{\begingroup\@sanitize@url \@url }%
\providecommand \@url [1]{\endgroup\@href {#1}{\urlprefix }}%
\providecommand \urlprefix  [0]{URL }%
\providecommand \Eprint [0]{\href }%
\providecommand \doibase [0]{https://doi.org/}%
\providecommand \selectlanguage [0]{\@gobble}%
\providecommand \bibinfo  [0]{\@secondoftwo}%
\providecommand \bibfield  [0]{\@secondoftwo}%
\providecommand \translation [1]{[#1]}%
\providecommand \BibitemOpen [0]{}%
\providecommand \bibitemStop [0]{}%
\providecommand \bibitemNoStop [0]{.\EOS\space}%
\providecommand \EOS [0]{\spacefactor3000\relax}%
\providecommand \BibitemShut  [1]{\csname bibitem#1\endcsname}%
\let\auto@bib@innerbib\@empty
\bibitem [{\citenamefont {Lewenstein}\ \emph {et~al.}(2007)\citenamefont
  {Lewenstein}, \citenamefont {Sanpera}, \citenamefont {Ahufinger},
  \citenamefont {Damski}, \citenamefont {Sen},\ and\ \citenamefont
  {Sen}}]{Lewenstein2007}%
  \BibitemOpen
  \bibfield  {author} {\bibinfo {author} {\bibfnamefont {M.}~\bibnamefont
  {Lewenstein}}, \bibinfo {author} {\bibfnamefont {A.}~\bibnamefont {Sanpera}},
  \bibinfo {author} {\bibfnamefont {V.}~\bibnamefont {Ahufinger}}, \bibinfo
  {author} {\bibfnamefont {B.}~\bibnamefont {Damski}}, \bibinfo {author}
  {\bibfnamefont {A.}~\bibnamefont {Sen}},\ and\ \bibinfo {author}
  {\bibfnamefont {U.}~\bibnamefont {Sen}},\ }\bibfield  {title} {\bibinfo
  {title} {{Ultracold atomic gases in optical lattices: mimicking condensed
  matter physics and beyond}},\ }\href
  {https://doi.org/10.1080/00018730701223200} {\bibfield  {journal} {\bibinfo
  {journal} {Adv. Phys.}\ }\textbf {\bibinfo {volume} {56}},\ \bibinfo
  {pages} {243} (\bibinfo {year} {2007})}\BibitemShut {NoStop}%
\bibitem [{\citenamefont {Bloch}\ \emph {et~al.}(2008)\citenamefont {Bloch},
  \citenamefont {Dalibard},\ and\ \citenamefont {Zwerger}}]{Bloch2008}%
  \BibitemOpen
  \bibfield  {author} {\bibinfo {author} {\bibfnamefont {I.}~\bibnamefont
  {Bloch}}, \bibinfo {author} {\bibfnamefont {J.}~\bibnamefont {Dalibard}},\
  and\ \bibinfo {author} {\bibfnamefont {W.}~\bibnamefont {Zwerger}},\
  }\bibfield  {title} {\bibinfo {title} {{Many-body physics with ultracold
  gases}},\ }\href {https://doi.org/10.1103/RevModPhys.80.885} {\bibfield
  {journal} {\bibinfo  {journal} {Rev. Mod. Phys.}\ }\textbf
  {\bibinfo {volume} {80}},\ \bibinfo {pages} {885} (\bibinfo {year} {2008})}\BibitemShut
  {NoStop}%
\bibitem [{\citenamefont {Micheli}\ \emph {et~al.}(2006)\citenamefont
  {Micheli}, \citenamefont {Brennen},\ and\ \citenamefont
  {Zoller}}]{Micheli2006}%
  \BibitemOpen
  \bibfield  {author} {\bibinfo {author} {\bibfnamefont {A.}~\bibnamefont
  {Micheli}}, \bibinfo {author} {\bibfnamefont {G.~K.}\ \bibnamefont
  {Brennen}},\ and\ \bibinfo {author} {\bibfnamefont {P.}~\bibnamefont
  {Zoller}},\ }\bibfield  {title} {\bibinfo {title} {{A toolbox for
  lattice-spin models with polar molecules}},\ }\href
  {https://doi.org/10.1038/nphys287} {\bibfield  {journal} {\bibinfo  {journal}
  {Nat. Phys.}\ }\textbf {\bibinfo {volume} {2}},\ \bibinfo {pages} {341}
  (\bibinfo {year} {2006})}\BibitemShut {NoStop}%
\bibitem [{\citenamefont {Chin}\ \emph {et~al.}(2009)\citenamefont {Chin},
  \citenamefont {Flambaum},\ and\ \citenamefont {Kozlov}}]{Chin2009}%
  \BibitemOpen
  \bibfield  {author} {\bibinfo {author} {\bibfnamefont {C.}~\bibnamefont
  {Chin}}, \bibinfo {author} {\bibfnamefont {V.~V.}\ \bibnamefont {Flambaum}},\
  and\ \bibinfo {author} {\bibfnamefont {M.~G.}\ \bibnamefont {Kozlov}},\
  }\bibfield  {title} {\bibinfo {title} {{Ultracold molecules: new probes on
  the variation of fundamental constants}},\ }\href
  {https://doi.org/10.1088/1367-2630/11/5/055048} {\bibfield  {journal}
  {\bibinfo  {journal} {New J. Phys.}\ }\textbf {\bibinfo {volume}
  {11}},\ \bibinfo {pages} {055048} (\bibinfo {year} {2009})}\BibitemShut
  {NoStop}%
\bibitem [{\citenamefont {Carr}\ \emph {et~al.}(2009)\citenamefont {Carr},
  \citenamefont {DeMille}, \citenamefont {Krems},\ and\ \citenamefont
  {Ye}}]{Carr2009}%
  \BibitemOpen
  \bibfield  {author} {\bibinfo {author} {\bibfnamefont {L.~D.}\ \bibnamefont
  {Carr}}, \bibinfo {author} {\bibfnamefont {D.}~\bibnamefont {DeMille}},
  \bibinfo {author} {\bibfnamefont {R.~V.}\ \bibnamefont {Krems}},\ and\
  \bibinfo {author} {\bibfnamefont {J.}~\bibnamefont {Ye}},\ }\bibfield
  {title} {\bibinfo {title} {{Cold and ultracold molecules: science, technology
  and applications}},\ }\href {https://doi.org/10.1088/1367-2630/11/5/055049}
  {\bibfield  {journal} {\bibinfo  {journal} {New J. Phys.}\ }\textbf
  {\bibinfo {volume} {11}},\ \bibinfo {pages} {055049} (\bibinfo {year}
  {2009})}\BibitemShut {NoStop}%
\bibitem [{\citenamefont {Jaksch}\ \emph {et~al.}(1998)\citenamefont {Jaksch},
  \citenamefont {Bruder}, \citenamefont {Cirac}, \citenamefont {Gardiner},\
  and\ \citenamefont {Zoller}}]{Jaksch1998}%
  \BibitemOpen
  \bibfield  {author} {\bibinfo {author} {\bibfnamefont {D.}~\bibnamefont
  {Jaksch}}, \bibinfo {author} {\bibfnamefont {C.}~\bibnamefont {Bruder}},
  \bibinfo {author} {\bibfnamefont {J.~I.}\ \bibnamefont {Cirac}}, \bibinfo
  {author} {\bibfnamefont {C.~W.}\ \bibnamefont {Gardiner}},\ and\ \bibinfo
  {author} {\bibfnamefont {P.}~\bibnamefont {Zoller}},\ }\bibfield  {title}
  {\bibinfo {title} {{Cold Bosonic Atoms in Optical Lattices}},\ }\href
  {https://doi.org/10.1103/PhysRevLett.81.3108} {\bibfield  {journal} {\bibinfo
   {journal} {Phys. Rev. Lett.}\ }\textbf {\bibinfo {volume} {81}},\
  \bibinfo {pages} {3108} (\bibinfo {year} {1998})}\BibitemShut {NoStop}%
\bibitem [{\citenamefont {Greiner}\ \emph
  {et~al.}(2002{\natexlab{a}})\citenamefont {Greiner}, \citenamefont {Mandel},
  \citenamefont {Esslinger}, \citenamefont {H{\"{a}}nsch},\ and\ \citenamefont
  {Bloch}}]{Greiner2002a}%
  \BibitemOpen
  \bibfield  {author} {\bibinfo {author} {\bibfnamefont {M.}~\bibnamefont
  {Greiner}}, \bibinfo {author} {\bibfnamefont {O.}~\bibnamefont {Mandel}},
  \bibinfo {author} {\bibfnamefont {T.}~\bibnamefont {Esslinger}}, \bibinfo
  {author} {\bibfnamefont {T.~W.}\ \bibnamefont {H{\"{a}}nsch}},\ and\ \bibinfo
  {author} {\bibfnamefont {I.}~\bibnamefont {Bloch}},\ }\bibfield  {title}
  {\bibinfo {title} {{Quantum phase transition from a superfluid to a Mott
  insulator in a gas of ultracold atoms}},\ }\href
  {https://doi.org/10.1038/415039a} {\bibfield  {journal} {\bibinfo  {journal}
  {Nature (London)}\ }\textbf {\bibinfo {volume} {415}},\ \bibinfo {pages} {39}
  (\bibinfo {year} {2002}{\natexlab{a}})}\BibitemShut {NoStop}%
\bibitem [{\citenamefont {St{\"{o}}ferle}\ \emph {et~al.}(2004)\citenamefont
  {St{\"{o}}ferle}, \citenamefont {Moritz}, \citenamefont {Schori},
  \citenamefont {K{\"{o}}hl},\ and\ \citenamefont {Esslinger}}]{Stoferle2004}%
  \BibitemOpen
  \bibfield  {author} {\bibinfo {author} {\bibfnamefont {T.}~\bibnamefont
  {St{\"{o}}ferle}}, \bibinfo {author} {\bibfnamefont {H.}~\bibnamefont
  {Moritz}}, \bibinfo {author} {\bibfnamefont {C.}~\bibnamefont {Schori}},
  \bibinfo {author} {\bibfnamefont {M.}~\bibnamefont {K{\"{o}}hl}},\ and\
  \bibinfo {author} {\bibfnamefont {T.}~\bibnamefont {Esslinger}},\ }\bibfield
  {title} {\bibinfo {title} {{Transition from a Strongly Interacting 1D
  Superfluid to a Mott Insulator}},\ }\href
  {https://doi.org/10.1103/PhysRevLett.92.130403} {\bibfield  {journal}
  {\bibinfo  {journal} {Phys. Rev. Lett.}\ }\textbf {\bibinfo {volume}
  {92}},\ \bibinfo {pages} {130403} (\bibinfo {year} {2004})}\BibitemShut
  {NoStop}%
\bibitem [{\citenamefont {Spielman}\ \emph {et~al.}(2007)\citenamefont
  {Spielman}, \citenamefont {Phillips},\ and\ \citenamefont
  {Porto}}]{Spielman2007}%
  \BibitemOpen
  \bibfield  {author} {\bibinfo {author} {\bibfnamefont {I.~B.}\ \bibnamefont
  {Spielman}}, \bibinfo {author} {\bibfnamefont {W.~D.}\ \bibnamefont
  {Phillips}},\ and\ \bibinfo {author} {\bibfnamefont {J.~V.}\ \bibnamefont
  {Porto}},\ }\bibfield  {title} {\bibinfo {title} {{Mott-Insulator Transition
  in a Two-Dimensional Atomic Bose Gas}},\ }\href
  {https://doi.org/10.1103/PhysRevLett.98.080404} {\bibfield  {journal}
  {\bibinfo  {journal} {Phys. Rev. Lett.}\ }\textbf {\bibinfo {volume}
  {98}},\ \bibinfo {pages} {080404} (\bibinfo {year} {2007})}\BibitemShut
  {NoStop}%
\bibitem [{\citenamefont {Spielman}\ \emph {et~al.}(2008)\citenamefont
  {Spielman}, \citenamefont {Phillips},\ and\ \citenamefont
  {Porto}}]{Spielman2008}%
  \BibitemOpen
  \bibfield  {author} {\bibinfo {author} {\bibfnamefont {I.~B.}\ \bibnamefont
  {Spielman}}, \bibinfo {author} {\bibfnamefont {W.~D.}\ \bibnamefont
  {Phillips}},\ and\ \bibinfo {author} {\bibfnamefont {J.~V.}\ \bibnamefont
  {Porto}},\ }\bibfield  {title} {\bibinfo {title} {{Condensate Fraction in a
  2D Bose Gas Measured across the Mott-Insulator Transition}},\ }\href
  {https://doi.org/10.1103/PhysRevLett.100.120402} {\bibfield  {journal}
  {\bibinfo  {journal} {Phys. Rev. Lett.}\ }\textbf {\bibinfo {volume}
  {100}},\ \bibinfo {pages} {120402} (\bibinfo {year} {2008})}\BibitemShut
  {NoStop}%
\bibitem [{\citenamefont {K{\"{o}}hl}\ \emph {et~al.}(2005)\citenamefont
  {K{\"{o}}hl}, \citenamefont {Moritz}, \citenamefont {St{\"{o}}ferle},
  \citenamefont {G{\"{u}}nter},\ and\ \citenamefont {Esslinger}}]{Kohl2005}%
  \BibitemOpen
  \bibfield  {author} {\bibinfo {author} {\bibfnamefont {M.}~\bibnamefont
  {K{\"{o}}hl}}, \bibinfo {author} {\bibfnamefont {H.}~\bibnamefont {Moritz}},
  \bibinfo {author} {\bibfnamefont {T.}~\bibnamefont {St{\"{o}}ferle}},
  \bibinfo {author} {\bibfnamefont {K.}~\bibnamefont {G{\"{u}}nter}},\ and\
  \bibinfo {author} {\bibfnamefont {T.}~\bibnamefont {Esslinger}},\ }\bibfield
  {title} {\bibinfo {title} {{Fermionic Atoms in a Three Dimensional Optical
  Lattice: Observing Fermi Surfaces, Dynamics, and Interactions}},\ }\href
  {https://doi.org/10.1103/PhysRevLett.94.080403} {\bibfield  {journal}
  {\bibinfo  {journal} {Phys. Rev. Lett.}\ }\textbf {\bibinfo {volume}
  {94}},\ \bibinfo {pages} {080403} (\bibinfo {year} {2005})}\BibitemShut
  {NoStop}%
\bibitem [{\citenamefont {Parsons}\ \emph {et~al.}(2016)\citenamefont
  {Parsons}, \citenamefont {Mazurenko}, \citenamefont {Chiu}, \citenamefont
  {Ji}, \citenamefont {Greif},\ and\ \citenamefont {Greiner}}]{Parsons2016}%
  \BibitemOpen
  \bibfield  {author} {\bibinfo {author} {\bibfnamefont {M.~F.}\ \bibnamefont
  {Parsons}}, \bibinfo {author} {\bibfnamefont {A.}~\bibnamefont {Mazurenko}},
  \bibinfo {author} {\bibfnamefont {C.~S.}\ \bibnamefont {Chiu}}, \bibinfo
  {author} {\bibfnamefont {G.}~\bibnamefont {Ji}}, \bibinfo {author}
  {\bibfnamefont {D.}~\bibnamefont {Greif}},\ and\ \bibinfo {author}
  {\bibfnamefont {M.}~\bibnamefont {Greiner}},\ }\bibfield  {title} {\bibinfo
  {title} {{Site-resolved measurement of the spin-correlation function in the
  Fermi-Hubbard model}},\ }\href {https://doi.org/10.1126/science.aag1430}
  {\bibfield  {journal} {\bibinfo  {journal} {Science}\ }\textbf {\bibinfo
  {volume} {353}},\ \bibinfo {pages} {1253} (\bibinfo {year} {2016})}\BibitemShut {NoStop}%
\bibitem [{\citenamefont {Esslinger}(2010)}]{Esslinger2010}%
  \BibitemOpen
  \bibfield  {author} {\bibinfo {author} {\bibfnamefont {T.}~\bibnamefont
  {Esslinger}},\ }\bibfield  {title} {\bibinfo {title} {{Fermi-Hubbard Physics
  with Atoms in an Optical Lattice}},\ }\href
  {https://doi.org/10.1146/annurev-conmatphys-070909-104059} {\bibfield
  {journal} {\bibinfo  {journal} {Annu. Rev. Condens. Matter Phys.}\
  }\textbf {\bibinfo {volume} {1}},\ \bibinfo {pages} {129} (\bibinfo {year}
  {2010})}\BibitemShut {NoStop}%
\bibitem [{\citenamefont {Wall}\ \emph {et~al.}(2015)\citenamefont {Wall},
  \citenamefont {Maeda},\ and\ \citenamefont {Carr}}]{Wall2015}%
  \BibitemOpen
  \bibfield  {author} {\bibinfo {author} {\bibfnamefont {M.~L.}\ \bibnamefont
  {Wall}}, \bibinfo {author} {\bibfnamefont {K.}~\bibnamefont {Maeda}},\ and\
  \bibinfo {author} {\bibfnamefont {L.~D.}\ \bibnamefont {Carr}},\ }\bibfield
  {title} {\bibinfo {title} {{Realizing unconventional quantum magnetism with
  symmetric top molecules}},\ }\href
  {https://doi.org/10.1088/1367-2630/17/2/025001} {\bibfield  {journal}
  {\bibinfo  {journal} {New J. Phys.}\ }\textbf {\bibinfo {volume}
  {17}},\ \bibinfo {pages} {025001} (\bibinfo {year} {2015})}\BibitemShut
  {NoStop}%
\bibitem [{\citenamefont {Pelegr{\'{i}}}\ \emph {et~al.}(2019)\citenamefont
  {Pelegr{\'{i}}}, \citenamefont {Mompart}, \citenamefont {Ahufinger},\ and\
  \citenamefont {Daley}}]{Pelegri2019}%
  \BibitemOpen
  \bibfield  {author} {\bibinfo {author} {\bibfnamefont {G.}~\bibnamefont
  {Pelegr{\'{i}}}}, \bibinfo {author} {\bibfnamefont {J.}~\bibnamefont
  {Mompart}}, \bibinfo {author} {\bibfnamefont {V.}~\bibnamefont {Ahufinger}},\
  and\ \bibinfo {author} {\bibfnamefont {A.~J.}\ \bibnamefont {Daley}},\
  }\bibfield  {title} {\bibinfo {title} {{Quantum magnetism with ultracold
  bosons carrying orbital angular momentum}},\ }\href
  {https://doi.org/10.1103/physreva.100.023615} {\bibfield  {journal} {\bibinfo
   {journal} {Phys. Rev. A}\ }\textbf {\bibinfo {volume} {100}},\ \bibinfo
  {pages} {023615} (\bibinfo {year} {2019})}\BibitemShut {NoStop}%
\bibitem [{\citenamefont {Simon}\ \emph {et~al.}(2011)\citenamefont {Simon},
  \citenamefont {Bakr}, \citenamefont {Ma}, \citenamefont {Tai}, \citenamefont
  {Preiss},\ and\ \citenamefont {Greiner}}]{Simon2011}%
  \BibitemOpen
  \bibfield  {author} {\bibinfo {author} {\bibfnamefont {J.}~\bibnamefont
  {Simon}}, \bibinfo {author} {\bibfnamefont {W.~S.}\ \bibnamefont {Bakr}},
  \bibinfo {author} {\bibfnamefont {R.}~\bibnamefont {Ma}}, \bibinfo {author}
  {\bibfnamefont {M.~E.}\ \bibnamefont {Tai}}, \bibinfo {author} {\bibfnamefont
  {P.~M.}\ \bibnamefont {Preiss}},\ and\ \bibinfo {author} {\bibfnamefont
  {M.}~\bibnamefont {Greiner}},\ }\bibfield  {title} {\bibinfo {title}
  {{Quantum simulation of antiferromagnetic spin chains in an optical
  lattice}},\ }\href {https://doi.org/10.1038/nature09994} {\bibfield
  {journal} {\bibinfo  {journal} {Nature (London)}\ }\textbf {\bibinfo {volume} {472}},\
  \bibinfo {pages} {307} (\bibinfo {year} {2011})}\BibitemShut {NoStop}%
\bibitem [{\citenamefont {Fukuhara}\ \emph
  {et~al.}(2013{\natexlab{a}})\citenamefont {Fukuhara}, \citenamefont
  {Schau{\ss}}, \citenamefont {Endres}, \citenamefont {Hild}, \citenamefont
  {Cheneau}, \citenamefont {Bloch},\ and\ \citenamefont
  {Gross}}]{Fukuhara2013b}%
  \BibitemOpen
  \bibfield  {author} {\bibinfo {author} {\bibfnamefont {T.}~\bibnamefont
  {Fukuhara}}, \bibinfo {author} {\bibfnamefont {P.}~\bibnamefont
  {Schau{\ss}}}, \bibinfo {author} {\bibfnamefont {M.}~\bibnamefont {Endres}},
  \bibinfo {author} {\bibfnamefont {S.}~\bibnamefont {Hild}}, \bibinfo {author}
  {\bibfnamefont {M.}~\bibnamefont {Cheneau}}, \bibinfo {author} {\bibfnamefont
  {I.}~\bibnamefont {Bloch}},\ and\ \bibinfo {author} {\bibfnamefont
  {C.}~\bibnamefont {Gross}},\ }\bibfield  {title} {\bibinfo {title}
  {{Microscopic observation of magnon bound states and their dynamics}},\
  }\href {https://doi.org/10.1038/nature12541} {\bibfield  {journal} {\bibinfo
  {journal} {Nature (London)}\ }\textbf {\bibinfo {volume} {502}},\ \bibinfo {pages}
  {76} (\bibinfo {year} {2013}{\natexlab{a}})}\BibitemShut {NoStop}%
\bibitem [{\citenamefont {Fukuhara}\ \emph {et~al.}(2015)\citenamefont
  {Fukuhara}, \citenamefont {Hild}, \citenamefont {Zeiher}, \citenamefont
  {Schau{\ss}}, \citenamefont {Bloch}, \citenamefont {Endres},\ and\
  \citenamefont {Gross}}]{Fukuhara2015}%
  \BibitemOpen
  \bibfield  {author} {\bibinfo {author} {\bibfnamefont {T.}~\bibnamefont
  {Fukuhara}}, \bibinfo {author} {\bibfnamefont {S.}~\bibnamefont {Hild}},
  \bibinfo {author} {\bibfnamefont {J.}~\bibnamefont {Zeiher}}, \bibinfo
  {author} {\bibfnamefont {P.}~\bibnamefont {Schau{\ss}}}, \bibinfo {author}
  {\bibfnamefont {I.}~\bibnamefont {Bloch}}, \bibinfo {author} {\bibfnamefont
  {M.}~\bibnamefont {Endres}},\ and\ \bibinfo {author} {\bibfnamefont
  {C.}~\bibnamefont {Gross}},\ }\bibfield  {title} {\bibinfo {title}
  {{Spatially Resolved Detection of a Spin-Entanglement Wave in a Bose-Hubbard
  Chain}},\ }\href {https://doi.org/10.1103/PhysRevLett.115.035302} {\bibfield
  {journal} {\bibinfo  {journal} {Phys. Rev. Lett.}\ }\textbf {\bibinfo
  {volume} {115}},\ \bibinfo {pages} {035302} (\bibinfo {year} {2015})}\BibitemShut
  {NoStop}%
\bibitem [{\citenamefont {Yan}\ \emph {et~al.}(2013)\citenamefont {Yan},
  \citenamefont {Moses}, \citenamefont {Gadway}, \citenamefont {Covey},
  \citenamefont {Hazzard}, \citenamefont {Rey}, \citenamefont {Jin},\ and\
  \citenamefont {Ye}}]{Yan2013}%
  \BibitemOpen
  \bibfield  {author} {\bibinfo {author} {\bibfnamefont {B.}~\bibnamefont
  {Yan}}, \bibinfo {author} {\bibfnamefont {S.~A.}\ \bibnamefont {Moses}},
  \bibinfo {author} {\bibfnamefont {B.}~\bibnamefont {Gadway}}, \bibinfo
  {author} {\bibfnamefont {J.~P.}\ \bibnamefont {Covey}}, \bibinfo {author}
  {\bibfnamefont {K.~R.}\ \bibnamefont {Hazzard}}, \bibinfo {author}
  {\bibfnamefont {A.~M.}\ \bibnamefont {Rey}}, \bibinfo {author} {\bibfnamefont
  {D.~S.}\ \bibnamefont {Jin}},\ and\ \bibinfo {author} {\bibfnamefont
  {J.}~\bibnamefont {Ye}},\ }\bibfield  {title} {\bibinfo {title} {{Observation
  of dipolar spin-exchange interactions with lattice-confined polar
  molecules}},\ }\href {https://doi.org/10.1038/nature12483} {\bibfield
  {journal} {\bibinfo  {journal} {Nature (London)}\ }\textbf {\bibinfo {volume} {501}},\
  \bibinfo {pages} {521} (\bibinfo {year} {2013})}\BibitemShut {NoStop}%
\bibitem [{\citenamefont {Hazzard}\ \emph {et~al.}(2014)\citenamefont
  {Hazzard}, \citenamefont {Gadway}, \citenamefont {Foss-Feig}, \citenamefont
  {Yan}, \citenamefont {Moses}, \citenamefont {Covey}, \citenamefont {Yao},
  \citenamefont {Lukin}, \citenamefont {Ye}, \citenamefont {Jin},\ and\
  \citenamefont {Rey}}]{Hazzard2014}%
  \BibitemOpen
  \bibfield  {author} {\bibinfo {author} {\bibfnamefont {K.~R.}\ \bibnamefont
  {Hazzard}}, \bibinfo {author} {\bibfnamefont {B.}~\bibnamefont {Gadway}},
  \bibinfo {author} {\bibfnamefont {M.}~\bibnamefont {Foss-Feig}}, \bibinfo
  {author} {\bibfnamefont {B.}~\bibnamefont {Yan}}, \bibinfo {author}
  {\bibfnamefont {S.~A.}\ \bibnamefont {Moses}}, \bibinfo {author}
  {\bibfnamefont {J.~P.}\ \bibnamefont {Covey}}, \bibinfo {author}
  {\bibfnamefont {N.~Y.}\ \bibnamefont {Yao}}, \bibinfo {author} {\bibfnamefont
  {M.~D.}\ \bibnamefont {Lukin}}, \bibinfo {author} {\bibfnamefont
  {J.}~\bibnamefont {Ye}}, \bibinfo {author} {\bibfnamefont {D.~S.}\
  \bibnamefont {Jin}},\ and\ \bibinfo {author} {\bibfnamefont {A.~M.}\
  \bibnamefont {Rey}},\ }\bibfield  {title} {\bibinfo {title} {{Many-Body
  Dynamics of Dipolar Molecules in an Optical Lattice}},\ }\href
  {https://doi.org/10.1103/PhysRevLett.113.195302} {\bibfield  {journal}
  {\bibinfo  {journal} {Phys. Rev. Lett.}\ }\textbf {\bibinfo {volume}
  {113}},\ \bibinfo {pages} {195302} (\bibinfo {year} {2014})}\BibitemShut {NoStop}%
\bibitem [{\citenamefont {Orioli}\ \emph {et~al.}(2018)\citenamefont {Orioli},
  \citenamefont {Signoles}, \citenamefont {Wildhagen}, \citenamefont
  {G{\"{u}}nter}, \citenamefont {Berges}, \citenamefont {Whitlock},\ and\
  \citenamefont {Weidem{\"{u}}ller}}]{Orioli2018}%
  \BibitemOpen
  \bibfield  {author} {\bibinfo {author} {\bibfnamefont {A.~P.}\ \bibnamefont
  {Orioli}}, \bibinfo {author} {\bibfnamefont {A.}~\bibnamefont {Signoles}},
  \bibinfo {author} {\bibfnamefont {H.}~\bibnamefont {Wildhagen}}, \bibinfo
  {author} {\bibfnamefont {G.}~\bibnamefont {G{\"{u}}nter}}, \bibinfo {author}
  {\bibfnamefont {J.}~\bibnamefont {Berges}}, \bibinfo {author} {\bibfnamefont
  {S.}~\bibnamefont {Whitlock}},\ and\ \bibinfo {author} {\bibfnamefont
  {M.}~\bibnamefont {Weidem{\"{u}}ller}},\ }\bibfield  {title} {\bibinfo
  {title} {{Relaxation of an Isolated Dipolar-Interacting Rydberg Quantum Spin
  System}},\ }\href {https://doi.org/10.1103/PhysRevLett.120.063601} {\bibfield
   {journal} {\bibinfo  {journal} {Phys. Rev. Lett.}\ }\textbf {\bibinfo
  {volume} {120}},\ \bibinfo {pages} {063601} (\bibinfo {year} {2018})}\BibitemShut
  {NoStop}%
\bibitem [{\citenamefont {Amico}\ \emph {et~al.}(2005)\citenamefont {Amico},
  \citenamefont {Osterloh},\ and\ \citenamefont {Cataliotti}}]{Amico2005}%
  \BibitemOpen
  \bibfield  {author} {\bibinfo {author} {\bibfnamefont {L.}~\bibnamefont
  {Amico}}, \bibinfo {author} {\bibfnamefont {A.}~\bibnamefont {Osterloh}},\
  and\ \bibinfo {author} {\bibfnamefont {F.}~\bibnamefont {Cataliotti}},\
  }\bibfield  {title} {\bibinfo {title} {{Quantum Many Particle Systems in
  Ring-Shaped Optical Lattices}},\ }\href
  {https://doi.org/10.1103/PhysRevLett.95.063201} {\bibfield  {journal}
  {\bibinfo  {journal} {Phys. Rev. Lett.}\ }\textbf {\bibinfo {volume}
  {95}},\ \bibinfo {pages} {063201} (\bibinfo {year} {2005})}\BibitemShut
  {NoStop}%
\bibitem [{\citenamefont {Henderson}\ \emph {et~al.}(2009)\citenamefont
  {Henderson}, \citenamefont {Ryu}, \citenamefont {MacCormick},\ and\
  \citenamefont {Boshier}}]{Henderson2009}%
  \BibitemOpen
  \bibfield  {author} {\bibinfo {author} {\bibfnamefont {K.}~\bibnamefont
  {Henderson}}, \bibinfo {author} {\bibfnamefont {C.}~\bibnamefont {Ryu}},
  \bibinfo {author} {\bibfnamefont {C.}~\bibnamefont {MacCormick}},\ and\
  \bibinfo {author} {\bibfnamefont {M.~G.}\ \bibnamefont {Boshier}},\
  }\bibfield  {title} {\bibinfo {title} {{Experimental demonstration of
  painting arbitrary and dynamic potentials for Bose-Einstein condensates}},\
  }\href {https://doi.org/10.1088/1367-2630/11/4/043030} {\bibfield  {journal}
  {\bibinfo  {journal} {New J. Phys.}\ }\textbf {\bibinfo {volume}
  {11}},\ \bibinfo {pages} {043030} (\bibinfo {year} {2009})}\BibitemShut {NoStop}%
\bibitem [{\citenamefont {Wirth}\ \emph {et~al.}(2011)\citenamefont {Wirth},
  \citenamefont {{\"{O}}lschl{\"{a}}ger},\ and\ \citenamefont
  {Hemmerich}}]{Wirth2011}%
  \BibitemOpen
  \bibfield  {author} {\bibinfo {author} {\bibfnamefont {G.}~\bibnamefont
  {Wirth}}, \bibinfo {author} {\bibfnamefont {M.}~\bibnamefont
  {{\"{O}}lschl{\"{a}}ger}},\ and\ \bibinfo {author} {\bibfnamefont
  {A.}~\bibnamefont {Hemmerich}},\ }\bibfield  {title} {\bibinfo {title}
  {{Evidence for orbital superfluidity in the P-band of a bipartite optical
  square lattice}},\ }\href {https://doi.org/10.1038/nphys1857} {\bibfield
  {journal} {\bibinfo  {journal} {Nat. Phys.}\ }\textbf {\bibinfo {volume}
  {7}},\ \bibinfo {pages} {147} (\bibinfo {year} {2011})}\BibitemShut {NoStop}%
\bibitem [{\citenamefont {Soltan-Panahi}\ \emph {et~al.}(2011)\citenamefont
  {Soltan-Panahi}, \citenamefont {Struck}, \citenamefont {Hauke}, \citenamefont
  {Bick}, \citenamefont {Plenkers}, \citenamefont {Meineke}, \citenamefont
  {Becker}, \citenamefont {Windpassinger}, \citenamefont {Lewenstein},\ and\
  \citenamefont {Sengstock}}]{Soltan-Panahi2011}%
  \BibitemOpen
  \bibfield  {author} {\bibinfo {author} {\bibfnamefont {P.}~\bibnamefont
  {Soltan-Panahi}}, \bibinfo {author} {\bibfnamefont {J.}~\bibnamefont
  {Struck}}, \bibinfo {author} {\bibfnamefont {P.}~\bibnamefont {Hauke}},
  \bibinfo {author} {\bibfnamefont {A.}~\bibnamefont {Bick}}, \bibinfo {author}
  {\bibfnamefont {W.}~\bibnamefont {Plenkers}}, \bibinfo {author}
  {\bibfnamefont {G.}~\bibnamefont {Meineke}}, \bibinfo {author} {\bibfnamefont
  {C.}~\bibnamefont {Becker}}, \bibinfo {author} {\bibfnamefont
  {P.}~\bibnamefont {Windpassinger}}, \bibinfo {author} {\bibfnamefont
  {M.}~\bibnamefont {Lewenstein}},\ and\ \bibinfo {author} {\bibfnamefont
  {K.}~\bibnamefont {Sengstock}},\ }\bibfield  {title} {\bibinfo {title}
  {{Multi-component quantum gases in spin-dependent hexagonal lattices}},\
  }\href {https://doi.org/10.1038/nphys1916} {\bibfield  {journal} {\bibinfo
  {journal} {Nat. Phys.}\ }\textbf {\bibinfo {volume} {7}},\ \bibinfo
  {pages} {434} (\bibinfo {year} {2011})}\BibitemShut {NoStop}%
\bibitem [{\citenamefont {Tarruell}\ \emph {et~al.}(2012)\citenamefont
  {Tarruell}, \citenamefont {Greif}, \citenamefont {Uehlinger}, \citenamefont
  {Jotzu},\ and\ \citenamefont {Esslinger}}]{Tarruell2012}%
  \BibitemOpen
  \bibfield  {author} {\bibinfo {author} {\bibfnamefont {L.}~\bibnamefont
  {Tarruell}}, \bibinfo {author} {\bibfnamefont {D.}~\bibnamefont {Greif}},
  \bibinfo {author} {\bibfnamefont {T.}~\bibnamefont {Uehlinger}}, \bibinfo
  {author} {\bibfnamefont {G.}~\bibnamefont {Jotzu}},\ and\ \bibinfo {author}
  {\bibfnamefont {T.}~\bibnamefont {Esslinger}},\ }\bibfield  {title} {\bibinfo
  {title} {{Creating, moving and merging Dirac points with a Fermi gas in a
  tunable honeycomb lattice}},\ }\href {https://doi.org/10.1038/nature10871}
  {\bibfield  {journal} {\bibinfo  {journal} {Nature (London)}\ }\textbf {\bibinfo
  {volume} {483}},\ \bibinfo {pages} {302} (\bibinfo {year} {2012})}\BibitemShut {NoStop}%
\bibitem [{\citenamefont {Jo}\ \emph {et~al.}(2012)\citenamefont {Jo},
  \citenamefont {Guzman}, \citenamefont {Thomas}, \citenamefont {Hosur},
  \citenamefont {Vishwanath},\ and\ \citenamefont {Stamper-Kurn}}]{Jo2012}%
  \BibitemOpen
  \bibfield  {author} {\bibinfo {author} {\bibfnamefont {G.~B.}\ \bibnamefont
  {Jo}}, \bibinfo {author} {\bibfnamefont {J.}~\bibnamefont {Guzman}}, \bibinfo
  {author} {\bibfnamefont {C.~K.}\ \bibnamefont {Thomas}}, \bibinfo {author}
  {\bibfnamefont {P.}~\bibnamefont {Hosur}}, \bibinfo {author} {\bibfnamefont
  {A.}~\bibnamefont {Vishwanath}},\ and\ \bibinfo {author} {\bibfnamefont
  {D.~M.}\ \bibnamefont {Stamper-Kurn}},\ }\bibfield  {title} {\bibinfo {title}
  {{Ultracold Atoms in a Tunable Optical Kagome Lattice}},\ }\href
  {https://doi.org/10.1103/PhysRevLett.108.045305} {\bibfield  {journal}
  {\bibinfo  {journal} {Phys. Rev. Lett.}\ }\textbf {\bibinfo {volume}
  {108}},\ \bibinfo {pages} {045305} (\bibinfo {year} {2012})}\BibitemShut {NoStop}%
\bibitem [{\citenamefont {Feshbach}(1958)}]{Feshbach1958}%
  \BibitemOpen
  \bibfield  {author} {\bibinfo {author} {\bibfnamefont {H.}~\bibnamefont
  {Feshbach}},\ }\bibfield  {title} {\bibinfo {title} {{Unified theory of
  nuclear reactions}},\ }\href {https://doi.org/10.1016/0003-4916(58)90007-1}
  {\bibfield  {journal} {\bibinfo  {journal} {Ann. Phys. (N.Y.)}\ }\textbf
  {\bibinfo {volume} {5}},\ \bibinfo {pages} {357} (\bibinfo {year}
  {1958})}\BibitemShut {NoStop}%
\bibitem [{\citenamefont {Fano}(1961)}]{Fano1961}%
  \BibitemOpen
  \bibfield  {author} {\bibinfo {author} {\bibfnamefont {U.}~\bibnamefont
  {Fano}},\ }\bibfield  {title} {\bibinfo {title} {{Effects of Configuration
  Interaction on Intensities and Phase Shifts}},\ }\href
  {https://doi.org/10.1103/PhysRev.124.1866} {\bibfield  {journal} {\bibinfo
  {journal} {Phys. Rev.}\ }\textbf {\bibinfo {volume} {124}},\ \bibinfo
  {pages} {1866} (\bibinfo {year} {1961})}\BibitemShut {NoStop}%
\bibitem [{\citenamefont {Tiesinga}\ \emph {et~al.}(1993)\citenamefont
  {Tiesinga}, \citenamefont {Verhaar},\ and\ \citenamefont
  {Stoof}}]{Tiesinga1993}%
  \BibitemOpen
  \bibfield  {author} {\bibinfo {author} {\bibfnamefont {E.}~\bibnamefont
  {Tiesinga}}, \bibinfo {author} {\bibfnamefont {B.~J.}\ \bibnamefont
  {Verhaar}},\ and\ \bibinfo {author} {\bibfnamefont {H.~T.~C.}\ \bibnamefont
  {Stoof}},\ }\bibfield  {title} {\bibinfo {title} {{Threshold and resonance
  phenomena in ultracold ground-state collisions}},\ }\href
  {https://doi.org/10.1103/PhysRevA.47.4114} {\bibfield  {journal} {\bibinfo
  {journal} {Phys. Rev. A}\ }\textbf {\bibinfo {volume} {47}},\ \bibinfo
  {pages} {4114} (\bibinfo {year} {1993})}\BibitemShut {NoStop}%
\bibitem [{\citenamefont {Stoof}\ \emph {et~al.}(1996)\citenamefont {Stoof},
  \citenamefont {Houbiers}, \citenamefont {Sackett},\ and\ \citenamefont
  {Hulet}}]{Stoof1996}%
  \BibitemOpen
  \bibfield  {author} {\bibinfo {author} {\bibfnamefont {H.~T.}\ \bibnamefont
  {Stoof}}, \bibinfo {author} {\bibfnamefont {M.}~\bibnamefont {Houbiers}},
  \bibinfo {author} {\bibfnamefont {C.~A.}\ \bibnamefont {Sackett}},\ and\
  \bibinfo {author} {\bibfnamefont {R.~G.}\ \bibnamefont {Hulet}},\ }\bibfield
  {title} {\bibinfo {title} {{Superfluidity of Spin-Polarized ${}^{6}$Li}},\ }\href
  {https://doi.org/10.1103/PhysRevLett.76.10} {\bibfield  {journal} {\bibinfo
  {journal} {Phys. Rev. Lett.}\ }\textbf {\bibinfo {volume} {76}},\
  \bibinfo {pages} {10} (\bibinfo {year} {1996})}\BibitemShut {NoStop}%
\bibitem [{\citenamefont {Greiner}\ \emph
  {et~al.}(2002{\natexlab{b}})\citenamefont {Greiner}, \citenamefont {Mandel},
  \citenamefont {H{\"{a}}nsch},\ and\ \citenamefont {Bloch}}]{Greiner2002b}%
  \BibitemOpen
  \bibfield  {author} {\bibinfo {author} {\bibfnamefont {M.}~\bibnamefont
  {Greiner}}, \bibinfo {author} {\bibfnamefont {O.}~\bibnamefont {Mandel}},
  \bibinfo {author} {\bibfnamefont {T.~W.}\ \bibnamefont {H{\"{a}}nsch}},\ and\
  \bibinfo {author} {\bibfnamefont {I.}~\bibnamefont {Bloch}},\ }\bibfield
  {title} {\bibinfo {title} {{Collapse and revival of the matter wave field of
  a Bose–Einstein condensate}},\ }\href {https://doi.org/10.1038/nature00968}
  {\bibfield  {journal} {\bibinfo  {journal} {Nature (London)}\ }\textbf {\bibinfo
  {volume} {419}},\ \bibinfo {pages} {51} (\bibinfo {year}
  {2002}{\natexlab{b}})}\BibitemShut {NoStop}%
\bibitem [{\citenamefont {Sadler}\ \emph {et~al.}(2006)\citenamefont {Sadler},
  \citenamefont {Higbie}, \citenamefont {Leslie}, \citenamefont
  {Vengalattore},\ and\ \citenamefont {Stamper-Kurn}}]{Sadler2006}%
  \BibitemOpen
  \bibfield  {author} {\bibinfo {author} {\bibfnamefont {L.~E.}\ \bibnamefont
  {Sadler}}, \bibinfo {author} {\bibfnamefont {J.~M.}\ \bibnamefont {Higbie}},
  \bibinfo {author} {\bibfnamefont {S.~R.}\ \bibnamefont {Leslie}}, \bibinfo
  {author} {\bibfnamefont {M.}~\bibnamefont {Vengalattore}},\ and\ \bibinfo
  {author} {\bibfnamefont {D.~M.}\ \bibnamefont {Stamper-Kurn}},\ }\bibfield
  {title} {\bibinfo {title} {{Spontaneous symmetry breaking in a quenched
  ferromagnetic spinor Bose-Einstein condensate}},\ }\href
  {https://doi.org/10.1038/nature05094} {\bibfield  {journal} {\bibinfo
  {journal} {Nature (London)}\ }\textbf {\bibinfo {volume} {443}},\ \bibinfo {pages}
  {312} (\bibinfo {year} {2006})}\BibitemShut {NoStop}%
\bibitem [{\citenamefont {Meinert}\ \emph {et~al.}(2013)\citenamefont
  {Meinert}, \citenamefont {Mark}, \citenamefont {Kirilov}, \citenamefont
  {Lauber}, \citenamefont {Weinmann}, \citenamefont {Daley},\ and\
  \citenamefont {N{\"{a}}gerl}}]{Meinert2013}%
  \BibitemOpen
  \bibfield  {author} {\bibinfo {author} {\bibfnamefont {F.}~\bibnamefont
  {Meinert}}, \bibinfo {author} {\bibfnamefont {M.~J.}\ \bibnamefont {Mark}},
  \bibinfo {author} {\bibfnamefont {E.}~\bibnamefont {Kirilov}}, \bibinfo
  {author} {\bibfnamefont {K.}~\bibnamefont {Lauber}}, \bibinfo {author}
  {\bibfnamefont {P.}~\bibnamefont {Weinmann}}, \bibinfo {author}
  {\bibfnamefont {A.~J.}\ \bibnamefont {Daley}},\ and\ \bibinfo {author}
  {\bibfnamefont {H.~C.}\ \bibnamefont {N{\"{a}}gerl}},\ }\bibfield  {title}
  {\bibinfo {title} {{Quantum Quench in an Atomic One-Dimensional Ising
  Chain}},\ }\href {https://doi.org/10.1103/PhysRevLett.111.053003} {\bibfield
  {journal} {\bibinfo  {journal} {Phys. Rev. Lett.}\ }\textbf {\bibinfo
  {volume} {111}},\ \bibinfo {pages} {053003} (\bibinfo {year}
  {2013})}\BibitemShut {NoStop}%
\bibitem [{\citenamefont {Lamporesi}\ \emph {et~al.}(2013)\citenamefont
  {Lamporesi}, \citenamefont {Donadello}, \citenamefont {Serafini},
  \citenamefont {Dalfovo},\ and\ \citenamefont {Ferrari}}]{Lamporesi2013}%
  \BibitemOpen
  \bibfield  {author} {\bibinfo {author} {\bibfnamefont {G.}~\bibnamefont
  {Lamporesi}}, \bibinfo {author} {\bibfnamefont {S.}~\bibnamefont
  {Donadello}}, \bibinfo {author} {\bibfnamefont {S.}~\bibnamefont {Serafini}},
  \bibinfo {author} {\bibfnamefont {F.}~\bibnamefont {Dalfovo}},\ and\ \bibinfo
  {author} {\bibfnamefont {G.}~\bibnamefont {Ferrari}},\ }\bibfield  {title}
  {\bibinfo {title} {{Spontaneous creation of Kibble-Zurek solitons in a
  Bose-Einstein condensate}},\ }\href {https://doi.org/10.1038/nphys2734}
  {\bibfield  {journal} {\bibinfo  {journal} {Nat. Phys.}\ }\textbf
  {\bibinfo {volume} {9}},\ \bibinfo {pages} {656} (\bibinfo {year} {2013})}\BibitemShut
  {NoStop}%
\bibitem [{\citenamefont {Hung}\ \emph {et~al.}(2013)\citenamefont {Hung},
  \citenamefont {Gurarie},\ and\ \citenamefont {Chin}}]{Hung2013}%
  \BibitemOpen
  \bibfield  {author} {\bibinfo {author} {\bibfnamefont {C.~L.}\ \bibnamefont
  {Hung}}, \bibinfo {author} {\bibfnamefont {V.}~\bibnamefont {Gurarie}},\ and\
  \bibinfo {author} {\bibfnamefont {C.}~\bibnamefont {Chin}},\ }\bibfield
  {title} {\bibinfo {title} {{From Cosmology to Cold Atoms: Observation of
  Sakharov Oscillations in a Quenched Atomic Superfluid}},\ }\href
  {https://doi.org/10.1126/science.1237557} {\bibfield  {journal} {\bibinfo
  {journal} {Science}\ }\textbf {\bibinfo {volume} {341}},\ \bibinfo {pages}
  {1213} (\bibinfo {year} {2013})}\BibitemShut {NoStop}%
\bibitem [{\citenamefont {Fukuhara}\ \emph
  {et~al.}(2013{\natexlab{b}})\citenamefont {Fukuhara}, \citenamefont
  {Kantian}, \citenamefont {Endres}, \citenamefont {Cheneau}, \citenamefont
  {Schau{\ss}}, \citenamefont {Hild}, \citenamefont {Bellem}, \citenamefont
  {Schollw{\"{o}}ck}, \citenamefont {Giamarchi}, \citenamefont {Gross},
  \citenamefont {Bloch},\ and\ \citenamefont {Kuhr}}]{Fukuhara2013a}%
  \BibitemOpen
  \bibfield  {author} {\bibinfo {author} {\bibfnamefont {T.}~\bibnamefont
  {Fukuhara}}, \bibinfo {author} {\bibfnamefont {A.}~\bibnamefont {Kantian}},
  \bibinfo {author} {\bibfnamefont {M.}~\bibnamefont {Endres}}, \bibinfo
  {author} {\bibfnamefont {M.}~\bibnamefont {Cheneau}}, \bibinfo {author}
  {\bibfnamefont {P.}~\bibnamefont {Schau{\ss}}}, \bibinfo {author}
  {\bibfnamefont {S.}~\bibnamefont {Hild}}, \bibinfo {author} {\bibfnamefont
  {D.}~\bibnamefont {Bellem}}, \bibinfo {author} {\bibfnamefont
  {U.}~\bibnamefont {Schollw{\"{o}}ck}}, \bibinfo {author} {\bibfnamefont
  {T.}~\bibnamefont {Giamarchi}}, \bibinfo {author} {\bibfnamefont
  {C.}~\bibnamefont {Gross}}, \bibinfo {author} {\bibfnamefont
  {I.}~\bibnamefont {Bloch}},\ and\ \bibinfo {author} {\bibfnamefont
  {S.}~\bibnamefont {Kuhr}},\ }\bibfield  {title} {\bibinfo {title} {{Quantum
  dynamics of a mobile spin impurity}},\ }\href
  {https://doi.org/10.1038/nphys2561} {\bibfield  {journal} {\bibinfo
  {journal} {Nat. Phys.}\ }\textbf {\bibinfo {volume} {9}},\ \bibinfo
  {pages} {235} (\bibinfo {year} {2013}{\natexlab{b}})}\BibitemShut {NoStop}%
\bibitem [{\citenamefont {Hild}\ \emph {et~al.}(2014)\citenamefont {Hild},
  \citenamefont {Fukuhara}, \citenamefont {Schau{\ss}}, \citenamefont {Zeiher},
  \citenamefont {Knap}, \citenamefont {Demler}, \citenamefont {Bloch},\ and\
  \citenamefont {Gross}}]{Hild2014}%
  \BibitemOpen
  \bibfield  {author} {\bibinfo {author} {\bibfnamefont {S.}~\bibnamefont
  {Hild}}, \bibinfo {author} {\bibfnamefont {T.}~\bibnamefont {Fukuhara}},
  \bibinfo {author} {\bibfnamefont {P.}~\bibnamefont {Schau{\ss}}}, \bibinfo
  {author} {\bibfnamefont {J.}~\bibnamefont {Zeiher}}, \bibinfo {author}
  {\bibfnamefont {M.}~\bibnamefont {Knap}}, \bibinfo {author} {\bibfnamefont
  {E.}~\bibnamefont {Demler}}, \bibinfo {author} {\bibfnamefont
  {I.}~\bibnamefont {Bloch}},\ and\ \bibinfo {author} {\bibfnamefont
  {C.}~\bibnamefont {Gross}},\ }\bibfield  {title} {\bibinfo {title}
  {{Far-from-Equilibrium Spin Transport in Heisenberg Quantum Magnets}},\
  }\href {https://doi.org/10.1103/PhysRevLett.113.147205} {\bibfield  {journal}
  {\bibinfo  {journal} {Phys. Rev. Lett.}\ }\textbf {\bibinfo {volume}
  {113}},\ \bibinfo {pages} {147205} (\bibinfo {year} {2014})}\BibitemShut
  {NoStop}%
\bibitem [{\citenamefont {Trotzky}\ \emph {et~al.}(2012)\citenamefont
  {Trotzky}, \citenamefont {Chen}, \citenamefont {Flesch}, \citenamefont
  {McCulloch}, \citenamefont {Schollw{\"{o}}ck}, \citenamefont {Eisert},\ and\
  \citenamefont {Bloch}}]{Trotzky2012}%
  \BibitemOpen
  \bibfield  {author} {\bibinfo {author} {\bibfnamefont {S.}~\bibnamefont
  {Trotzky}}, \bibinfo {author} {\bibfnamefont {Y.~A.}\ \bibnamefont {Chen}},
  \bibinfo {author} {\bibfnamefont {A.}~\bibnamefont {Flesch}}, \bibinfo
  {author} {\bibfnamefont {I.~P.}\ \bibnamefont {McCulloch}}, \bibinfo {author}
  {\bibfnamefont {U.}~\bibnamefont {Schollw{\"{o}}ck}}, \bibinfo {author}
  {\bibfnamefont {J.}~\bibnamefont {Eisert}},\ and\ \bibinfo {author}
  {\bibfnamefont {I.}~\bibnamefont {Bloch}},\ }\bibfield  {title} {\bibinfo
  {title} {{Probing the relaxation towards equilibrium in an isolated strongly
  correlated one-dimensional Bose gas}},\ }\href
  {https://doi.org/10.1038/nphys2232} {\bibfield  {journal} {\bibinfo
  {journal} {Nat. Phys.}\ }\textbf {\bibinfo {volume} {8}},\ \bibinfo
  {pages} {325} (\bibinfo {year} {2012})}\BibitemShut {NoStop}%
\bibitem [{\citenamefont {Kinoshita}\ \emph {et~al.}(2006)\citenamefont
  {Kinoshita}, \citenamefont {Wenger},\ and\ \citenamefont
  {Weiss}}]{Kinoshita2006}%
  \BibitemOpen
  \bibfield  {author} {\bibinfo {author} {\bibfnamefont {T.}~\bibnamefont
  {Kinoshita}}, \bibinfo {author} {\bibfnamefont {T.}~\bibnamefont {Wenger}},\
  and\ \bibinfo {author} {\bibfnamefont {D.~S.}\ \bibnamefont {Weiss}},\
  }\bibfield  {title} {\bibinfo {title} {{A quantum Newton's cradle}},\ }\href
  {https://doi.org/10.1038/nature04693} {\bibfield  {journal} {\bibinfo
  {journal} {Nature (London)}\ }\textbf {\bibinfo {volume} {440}},\ \bibinfo {pages}
  {900} (\bibinfo {year} {2006})}\BibitemShut {NoStop}%
\bibitem [{\citenamefont {Gring}\ \emph {et~al.}(2012)\citenamefont {Gring},
  \citenamefont {Kuhnert}, \citenamefont {Langen}, \citenamefont {Kitagawa},
  \citenamefont {Rauer}, \citenamefont {Schreitl}, \citenamefont {Mazets},
  \citenamefont {Smith}, \citenamefont {Demler},\ and\ \citenamefont
  {Schmiedmayer}}]{Gring2012}%
  \BibitemOpen
  \bibfield  {author} {\bibinfo {author} {\bibfnamefont {M.}~\bibnamefont
  {Gring}}, \bibinfo {author} {\bibfnamefont {M.}~\bibnamefont {Kuhnert}},
  \bibinfo {author} {\bibfnamefont {T.}~\bibnamefont {Langen}}, \bibinfo
  {author} {\bibfnamefont {T.}~\bibnamefont {Kitagawa}}, \bibinfo {author}
  {\bibfnamefont {B.}~\bibnamefont {Rauer}}, \bibinfo {author} {\bibfnamefont
  {M.}~\bibnamefont {Schreitl}}, \bibinfo {author} {\bibfnamefont
  {I.}~\bibnamefont {Mazets}}, \bibinfo {author} {\bibfnamefont {D.~A.}\
  \bibnamefont {Smith}}, \bibinfo {author} {\bibfnamefont {E.}~\bibnamefont
  {Demler}},\ and\ \bibinfo {author} {\bibfnamefont {J.}~\bibnamefont
  {Schmiedmayer}},\ }\bibfield  {title} {\bibinfo {title} {{Relaxation and
  Prethermalization in an Isolated Quantum System}},\ }\href@noop {} {\bibfield
   {journal} {\bibinfo  {journal} {Science}\ }\textbf {\bibinfo {volume}
  {337}},\ \bibinfo {pages} {1318} (\bibinfo {year} {2012})}\BibitemShut
  {NoStop}%
\bibitem [{\citenamefont {Tasaki}(1998)}]{Tasaki1998}%
  \BibitemOpen
  \bibfield  {author} {\bibinfo {author} {\bibfnamefont {H.}~\bibnamefont
  {Tasaki}},\ }\bibfield  {title} {\bibinfo {title} {{From Quantum Dynamics to
  the Canonical Distribution: General Picture and a Rigorous Example}},\ }\href
  {https://doi.org/10.1103/PhysRevLett.80.1373} {\bibfield  {journal} {\bibinfo
   {journal} {Phys. Rev. Lett.}\ }\textbf {\bibinfo {volume} {80}},\
  \bibinfo {pages} {1373} (\bibinfo {year} {1998})}\BibitemShut
  {NoStop}%
\bibitem [{\citenamefont {Reimann}(2008)}]{Reimann2008}%
  \BibitemOpen
  \bibfield  {author} {\bibinfo {author} {\bibfnamefont {P.}~\bibnamefont
  {Reimann}},\ }\bibfield  {title} {\bibinfo {title} {{Foundation of
  Statistical Mechanics under Experimentally Realistic Conditions}},\ }\href
  {https://doi.org/10.1103/PhysRevLett.101.190403} {\bibfield  {journal}
  {\bibinfo  {journal} {Phys. Rev. Lett.}\ }\textbf {\bibinfo {volume}
  {101}},\ \bibinfo {pages} {190403} (\bibinfo {year} {2008})}\BibitemShut {NoStop}%
\bibitem [{\citenamefont {Mori}\ \emph {et~al.}(2018)\citenamefont {Mori},
  \citenamefont {Ikeda}, \citenamefont {Kaminishi},\ and\ \citenamefont
  {Ueda}}]{Mori2018}%
  \BibitemOpen
  \bibfield  {author} {\bibinfo {author} {\bibfnamefont {T.}~\bibnamefont
  {Mori}}, \bibinfo {author} {\bibfnamefont {T.~N.}\ \bibnamefont {Ikeda}},
  \bibinfo {author} {\bibfnamefont {E.}~\bibnamefont {Kaminishi}},\ and\
  \bibinfo {author} {\bibfnamefont {M.}~\bibnamefont {Ueda}},\ }\bibfield
  {title} {\bibinfo {title} {{Thermalization and prethermalization in isolated
  quantum systems: a theoretical overview}},\ }\href
  {https://doi.org/10.1088/1361-6455/aabcdf} {\bibfield  {journal} {\bibinfo
  {journal} {J. Phys. B}\
  }\textbf {\bibinfo {volume} {51}},\ \bibinfo {pages} {112001} (\bibinfo
  {year} {2018})}\BibitemShut {NoStop}%
\bibitem [{\citenamefont {von Neumann}(2010)}]{VonNeumann2010}%
  \BibitemOpen
  \bibfield  {author} {\bibinfo {author} {\bibfnamefont {J.}~\bibnamefont {von
  Neumann}},\ }\bibfield  {title} {\bibinfo {title} {{Proof of the ergodic
  theorem and the H-theorem in quantum mechanics}},\ }\href
  {https://doi.org/10.1140/epjh/e2010-00008-5} {\bibfield  {journal} {\bibinfo
  {journal} {Eur. Phys. J. H}\ }\textbf {\bibinfo {volume}
  {35}},\ \bibinfo {pages} {201} (\bibinfo {year} {2010})}\BibitemShut
  {NoStop}%
\bibitem [{\citenamefont {Deutsch}(1991)}]{Deutsch1991}%
  \BibitemOpen
  \bibfield  {author} {\bibinfo {author} {\bibfnamefont {J.~M.}\ \bibnamefont
  {Deutsch}},\ }\bibfield  {title} {\bibinfo {title} {{Quantum statistical
  mechanics in a closed system}},\ }\href
  {https://doi.org/10.1103/PhysRevA.43.2046} {\bibfield  {journal} {\bibinfo
  {journal} {Phys. Rev. A}\ }\textbf {\bibinfo {volume} {43}},\ \bibinfo
  {pages} {2046} (\bibinfo {year} {1991})}\BibitemShut {NoStop}%
\bibitem [{\citenamefont {Srednicki}(1994)}]{Srednicki1994}%
  \BibitemOpen
  \bibfield  {author} {\bibinfo {author} {\bibfnamefont {M.}~\bibnamefont
  {Srednicki}},\ }\bibfield  {title} {\bibinfo {title} {{Chaos and quantum
  thermalization}},\ }\href {https://doi.org/10.1103/PhysRevE.50.888}
  {\bibfield  {journal} {\bibinfo  {journal} {Phys. Rev. E}\ }\textbf
  {\bibinfo {volume} {50}},\ \bibinfo {pages} {888} (\bibinfo {year}
  {1994})}\BibitemShut {NoStop}%
\bibitem [{\citenamefont {Rigol}\ \emph {et~al.}(2008)\citenamefont {Rigol},
  \citenamefont {Dunjko},\ and\ \citenamefont {Olshanii}}]{Rigol2008}%
  \BibitemOpen
  \bibfield  {author} {\bibinfo {author} {\bibfnamefont {M.}~\bibnamefont
  {Rigol}}, \bibinfo {author} {\bibfnamefont {V.}~\bibnamefont {Dunjko}},\ and\
  \bibinfo {author} {\bibfnamefont {M.}~\bibnamefont {Olshanii}},\ }\bibfield
  {title} {\bibinfo {title} {{Thermalization and its mechanism for generic
  isolated quantum systems}},\ }\href {https://doi.org/10.1038/nature06838}
  {\bibfield  {journal} {\bibinfo  {journal} {Nature (London)}\ }\textbf {\bibinfo
  {volume} {452}},\ \bibinfo {pages} {854} (\bibinfo {year}
  {2008})}\BibitemShut {NoStop}%
\bibitem [{\citenamefont {Srednicki}(1999)}]{Srednicki1999}%
  \BibitemOpen
  \bibfield  {author} {\bibinfo {author} {\bibfnamefont {M.}~\bibnamefont
  {Srednicki}},\ }\bibfield  {title} {\bibinfo {title} {{The approach to
  thermal equilibrium in quantized chaotic systems}},\ }\href@noop {}
  {\bibfield  {journal} {\bibinfo  {journal} {J. Phys. A}\ }\textbf {\bibinfo {volume} {32}},\ \bibinfo
  {pages} {1163} (\bibinfo {year} {1999})}\BibitemShut {NoStop}%
\bibitem [{\citenamefont {D'Alessio}\ \emph {et~al.}(2016)\citenamefont
  {D'Alessio}, \citenamefont {Kafri}, \citenamefont {Polkovnikov},\ and\
  \citenamefont {Rigol}}]{DAlessio2016}%
  \BibitemOpen
  \bibfield  {author} {\bibinfo {author} {\bibfnamefont {L.}~\bibnamefont
  {D'Alessio}}, \bibinfo {author} {\bibfnamefont {Y.}~\bibnamefont {Kafri}},
  \bibinfo {author} {\bibfnamefont {A.}~\bibnamefont {Polkovnikov}},\ and\
  \bibinfo {author} {\bibfnamefont {M.}~\bibnamefont {Rigol}},\ }\bibfield
  {title} {\bibinfo {title} {{From quantum chaos and eigenstate thermalization
  to statistical mechanics and thermodynamics}},\ }\href
  {https://doi.org/10.1080/00018732.2016.1198134} {\bibfield  {journal}
  {\bibinfo  {journal} {Adv. Phys.}\ }\textbf {\bibinfo {volume}
  {65}},\ \bibinfo {pages} {239} (\bibinfo {year} {2016})}\BibitemShut
  {NoStop}%
\bibitem [{\citenamefont {Anza}\ \emph {et~al.}(2018)\citenamefont {Anza},
  \citenamefont {Gogolin},\ and\ \citenamefont {Huber}}]{Anza2018}%
  \BibitemOpen
  \bibfield  {author} {\bibinfo {author} {\bibfnamefont {F.}~\bibnamefont
  {Anza}}, \bibinfo {author} {\bibfnamefont {C.}~\bibnamefont {Gogolin}},\ and\
  \bibinfo {author} {\bibfnamefont {M.}~\bibnamefont {Huber}},\ }\bibfield
  {title} {\bibinfo {title} {{Eigenstate Thermalization for Degenerate
  Observables}},\ }\href {https://doi.org/10.1103/PhysRevLett.120.150603}
  {\bibfield  {journal} {\bibinfo  {journal} {Phys. Rev. Lett.}\
  }\textbf {\bibinfo {volume} {120}},\ \bibinfo {pages} {150603} (\bibinfo
  {year} {2018})}\BibitemShut {NoStop}%
\bibitem [{\citenamefont {Goldstein}\ \emph {et~al.}(2017)\citenamefont
  {Goldstein}, \citenamefont {Huse}, \citenamefont {Lebowitz},\ and\
  \citenamefont {Tumulka}}]{Goldstein2017}%
  \BibitemOpen
  \bibfield  {author} {\bibinfo {author} {\bibfnamefont {S.}~\bibnamefont
  {Goldstein}}, \bibinfo {author} {\bibfnamefont {D.~A.}\ \bibnamefont {Huse}},
  \bibinfo {author} {\bibfnamefont {J.~L.}\ \bibnamefont {Lebowitz}},\ and\
  \bibinfo {author} {\bibfnamefont {R.}~\bibnamefont {Tumulka}},\ }\bibfield
  {title} {\bibinfo {title} {{Macroscopic and microscopic thermal
  equilibrium}},\ }\href {https://doi.org/10.1002/andp.201600301} {\bibfield
  {journal} {\bibinfo  {journal} {Ann. Phys. (Berlin)}\ }\textbf {\bibinfo
  {volume} {529}},\ \bibinfo {pages} {1600301} (\bibinfo {year}
  {2017})}\BibitemShut {NoStop}%
\bibitem [{\citenamefont {Biroli}\ \emph {et~al.}(2010)\citenamefont {Biroli},
  \citenamefont {Kollath},\ and\ \citenamefont {L{\"{a}}uchli}}]{Biroli2010}%
  \BibitemOpen
  \bibfield  {author} {\bibinfo {author} {\bibfnamefont {G.}~\bibnamefont
  {Biroli}}, \bibinfo {author} {\bibfnamefont {C.}~\bibnamefont {Kollath}},\
  and\ \bibinfo {author} {\bibfnamefont {A.~M.}\ \bibnamefont
  {L{\"{a}}uchli}},\ }\bibfield  {title} {\bibinfo {title} {{Effect of Rare
  Fluctuations on the Thermalization of Isolated Quantum Systems}},\ }\href
  {https://doi.org/10.1103/PhysRevLett.105.250401} {\bibfield  {journal}
  {\bibinfo  {journal} {Phys. Rev. Lett.}\ }\textbf {\bibinfo {volume}
  {105}},\ \bibinfo {pages} {250401} (\bibinfo {year} {2010})}\BibitemShut
  {NoStop}%
\bibitem [{\citenamefont {Iyoda}\ \emph {et~al.}(2017)\citenamefont {Iyoda},
  \citenamefont {Kaneko},\ and\ \citenamefont {Sagawa}}]{Iyoda2017}%
  \BibitemOpen
  \bibfield  {author} {\bibinfo {author} {\bibfnamefont {E.}~\bibnamefont
  {Iyoda}}, \bibinfo {author} {\bibfnamefont {K.}~\bibnamefont {Kaneko}},\ and\
  \bibinfo {author} {\bibfnamefont {T.}~\bibnamefont {Sagawa}},\ }\bibfield
  {title} {\bibinfo {title} {{Fluctuation Theorem for Many-Body Pure Quantum
  States}},\ }\href {https://doi.org/10.1103/PhysRevLett.119.100601} {\bibfield
   {journal} {\bibinfo  {journal} {Phys. Rev. Lett.}\ }\textbf {\bibinfo
  {volume} {119}},\ \bibinfo {pages} {100601} (\bibinfo {year}
  {2017})}\BibitemShut {NoStop}%
\bibitem [{\citenamefont {Mori}(2016)}]{Mori2016}%
  \BibitemOpen
  \bibfield  {author} {\bibinfo {author} {\bibfnamefont {T.}~\bibnamefont
  {Mori}},\ }\bibfield  {title} {\bibinfo {title} {{Weak eigenstate
  thermalization with large deviation bound}},\ }\href
  {http://arxiv.org/abs/1609.09776} {\bibfield  {journal}  }\Eprint
  {https://arxiv.org/abs/1609.09776} {arXiv:1609.09776} \BibitemShut {NoStop}%
\bibitem [{\citenamefont {Kuwahara}\ and\ \citenamefont
  {Saito}(2019)}]{Kuwahara2019}%
  \BibitemOpen
  \bibfield  {author} {\bibinfo {author} {\bibfnamefont {T.}~\bibnamefont
  {Kuwahara}}\ and\ \bibinfo {author} {\bibfnamefont {K.}~\bibnamefont
  {Saito}},\ }\bibfield  {title} {\bibinfo {title} {{Ensemble equivalence and
  eigenstate thermalization from clustering of correlation}},\ }\href
  {http://arxiv.org/abs/1905.01886} {\bibfield  {journal} } \Eprint
  {https://arxiv.org/abs/1905.01886} {arXiv:1905.01886} \BibitemShut {NoStop}%
\bibitem [{Note1()}]{Note1}%
  \BibitemOpen
  \bibinfo {note} {The initial state can be replaced with an appropriate pure
  quantum state that approximates the state in Ref.~\cite {Sugiura2013}, which
  gives the same results as the Gibbs state for both equilibrium \cite
  {Sugiura2013} and dynamical \cite {Shimizu2017,Endo2018}
  properties.}\BibitemShut {Stop}%
\bibitem [{\citenamefont {Sugiura}\ and\ \citenamefont
  {Shimizu}(2013)}]{Sugiura2013}%
  \BibitemOpen
  \bibfield  {author} {\bibinfo {author} {\bibfnamefont {S.}~\bibnamefont
  {Sugiura}}\ and\ \bibinfo {author} {\bibfnamefont {A.}~\bibnamefont
  {Shimizu}},\ }\bibfield  {title} {\bibinfo {title} {{Canonical Thermal Pure
  Quantum State}},\ }\href {https://doi.org/10.1103/PhysRevLett.111.010401}
  {\bibfield  {journal} {\bibinfo  {journal} {Phys. Rev. Lett.}\
  }\textbf {\bibinfo {volume} {111}},\ \bibinfo {pages} {010401} (\bibinfo
  {year} {2013})}\BibitemShut {NoStop}%
\bibitem [{\citenamefont {Shimizu}\ and\ \citenamefont
  {Fujikura}(2017)}]{Shimizu2017}%
  \BibitemOpen
  \bibfield  {author} {\bibinfo {author} {\bibfnamefont {A.}~\bibnamefont
  {Shimizu}}\ and\ \bibinfo {author} {\bibfnamefont {K.}~\bibnamefont
  {Fujikura}},\ }\bibfield  {title} {\bibinfo {title} {{Quantum violation of
  fluctuation-dissipation theorem}},\ }\href
  {https://doi.org/10.1088/1742-5468/aa5a67} {\bibfield  {journal} {\bibinfo
  {journal} {J. Stat. Mech.}\ }\textbf
  {\bibinfo {volume} } (\bibinfo {year}{2017}) \bibinfo {pages} {024004}}\BibitemShut {NoStop}%
\bibitem [{\citenamefont {Endo}\ \emph {et~al.}(2018)\citenamefont {Endo},
  \citenamefont {Hotta},\ and\ \citenamefont {Shimizu}}]{Endo2018}%
  \BibitemOpen
  \bibfield  {author} {\bibinfo {author} {\bibfnamefont {H.}~\bibnamefont
  {Endo}}, \bibinfo {author} {\bibfnamefont {C.}~\bibnamefont {Hotta}},\ and\
  \bibinfo {author} {\bibfnamefont {A.}~\bibnamefont {Shimizu}},\ }\bibfield
  {title} {\bibinfo {title} {{From Linear to Nonlinear Responses of Thermal
  Pure Quantum States}},\ }\href
  {https://doi.org/10.1103/PhysRevLett.121.220601} {\bibfield  {journal}
  {\bibinfo  {journal} {Phys. Rev. Lett.}\ }\textbf {\bibinfo {volume}
  {121}},\ \bibinfo {pages} {220601} (\bibinfo {year} {2018})}\BibitemShut {NoStop}%
\bibitem [{Note2()}]{Note2}%
  \BibitemOpen
  \bibinfo {note} {See Supplemental Material for detailed calculations, which
  includes Refs.~\cite {Mazur1969, Beugeling2014, Steinigeweg2014}}\BibitemShut
  {NoStop}%
\bibitem [{\citenamefont {Mazur}(1969)}]{Mazur1969}%
  \BibitemOpen
  \bibfield  {author} {\bibinfo {author} {\bibfnamefont {P.}~\bibnamefont
  {Mazur}},\ }\bibfield  {title} {\bibinfo {title} {{Non-ergodicity of phase
  functions in certain systems}},\ }\href
  {https://doi.org/10.1016/0031-8914(69)90185-2} {\bibfield  {journal}
  {\bibinfo  {journal} {Physica}\ }\textbf {\bibinfo {volume} {43}},\ \bibinfo
  {pages} {533} (\bibinfo {year} {1969})}\BibitemShut {NoStop}%
\bibitem [{\citenamefont {Beugeling}\ \emph {et~al.}(2014)\citenamefont
  {Beugeling}, \citenamefont {Moessner},\ and\ \citenamefont
  {Haque}}]{Beugeling2014}%
  \BibitemOpen
  \bibfield  {author} {\bibinfo {author} {\bibfnamefont {W.}~\bibnamefont
  {Beugeling}}, \bibinfo {author} {\bibfnamefont {R.}~\bibnamefont
  {Moessner}},\ and\ \bibinfo {author} {\bibfnamefont {M.}~\bibnamefont
  {Haque}},\ }\bibfield  {title} {\bibinfo {title} {{Finite-size scaling of
  eigenstate thermalization}},\ }\href
  {https://doi.org/10.1103/PhysRevE.89.042112} {\bibfield  {journal} {\bibinfo
  {journal} {Phys. Rev. E}\ }\textbf {\bibinfo {volume} {89}},\ \bibinfo {pages} {042112}
  (\bibinfo {year} {2014})}\BibitemShut {NoStop}%
\bibitem [{\citenamefont {Steinigeweg}\ \emph {et~al.}(2014)\citenamefont
  {Steinigeweg}, \citenamefont {Khodja}, \citenamefont {Niemeyer},
  \citenamefont {Gogolin},\ and\ \citenamefont {Gemmer}}]{Steinigeweg2014}%
  \BibitemOpen
  \bibfield  {author} {\bibinfo {author} {\bibfnamefont {R.}~\bibnamefont
  {Steinigeweg}}, \bibinfo {author} {\bibfnamefont {A.}~\bibnamefont {Khodja}},
  \bibinfo {author} {\bibfnamefont {H.}~\bibnamefont {Niemeyer}}, \bibinfo
  {author} {\bibfnamefont {C.}~\bibnamefont {Gogolin}},\ and\ \bibinfo {author}
  {\bibfnamefont {J.}~\bibnamefont {Gemmer}},\ }\bibfield  {title} {\bibinfo
  {title} {{Pushing the Limits of the Eigenstate Thermalization Hypothesis
  towards Mesoscopic Quantum Systems}},\ }\href
  {https://doi.org/10.1103/PhysRevLett.112.130403} {\bibfield  {journal}
  {\bibinfo  {journal} {Phys. Rev. Lett.}\ }\textbf {\bibinfo {volume}
  {112}},\ \bibinfo {pages} {130403} (\bibinfo {year} {2014})}\BibitemShut
  {NoStop}%
\bibitem [{Note3()}]{Note3}%
  \BibitemOpen
  \bibinfo {note} {The equality, $\chi _{N}^S(\vb *{k})=\chi _{N}^T(\vb *{k})$
  for all $\vb *{k}\not =\vb *{0}$ and for all $N$, can be proved only from the
  translation invariance \cite {Note2}.}\BibitemShut {Stop}%
\bibitem [{Note4()}]{Note4}%
  \BibitemOpen
  \bibinfo {note} {See Refs.~\cite {Lebowitz1967,Hansen2013} for related
  discussions.}\BibitemShut {Stop}%
\bibitem [{\citenamefont {Lebowitz}\ \emph {et~al.}(1967)\citenamefont
  {Lebowitz}, \citenamefont {Percus},\ and\ \citenamefont
  {Verlet}}]{Lebowitz1967}%
  \BibitemOpen
  \bibfield  {author} {\bibinfo {author} {\bibfnamefont {J.~L.}\ \bibnamefont
  {Lebowitz}}, \bibinfo {author} {\bibfnamefont {J.~K.}\ \bibnamefont
  {Percus}},\ and\ \bibinfo {author} {\bibfnamefont {L.}~\bibnamefont
  {Verlet}},\ }\bibfield  {title} {\bibinfo {title} {{Ensemble Dependence of
  Fluctuations with Application to Machine Computations}},\ }\href
  {https://doi.org/10.1103/PhysRev.153.250} {\bibfield  {journal} {\bibinfo
  {journal} {Phys. Rev.}\ }\textbf {\bibinfo {volume} {153}},\ \bibinfo
  {pages} {250} (\bibinfo {year} {1967})}\BibitemShut {NoStop}%
\bibitem [{\citenamefont {Hansen}\ and\ \citenamefont
  {McDonald}(2013)}]{Hansen2013}%
  \BibitemOpen
  \bibfield  {author} {\bibinfo {author} {\bibfnamefont {J.-P.}\ \bibnamefont
  {Hansen}}\ and\ \bibinfo {author} {\bibfnamefont {I.~R.}\ \bibnamefont
  {McDonald}},\ }\href@noop {} {\emph {\bibinfo {title} {Theory of Simple
  Liquids}}},\ \bibinfo {edition} {4th}\ ed.\ (\bibinfo  {publisher} {Academic
  Press, Oxford},\ \bibinfo {year} {2013})\BibitemShut {NoStop}%
\bibitem [{Note5()}]{Note5}%
  \BibitemOpen
  \bibinfo {note} {If $h=0$, so that $m_{\vb *{0}}=0$, we have $\chi _{\infty
  }^{S}(\vb *{0})=\chi _{\infty }^{T}(\vb *{0})$ unlike (\ref {chiT>chiS}).
  Even in such a case, Eq.~(\ref {eq:qch=T}) shows that $\chi _{\infty
  }^{\protect \text {qch}}(\vb *{k})$ is discontinuous at $\vb *k=\vb *0$
  unless the condition for (i) is satisfied.}\BibitemShut {Stop}%
\bibitem [{Note6()}]{Note6}%
  \BibitemOpen
  \bibinfo {note} {Note that condition~(\ref {eq:weakETHlike}) can be satisfied
  even when $C=0$, although $C$ do not vanish in our setting where $h\not =0$
  \cite {Note2}.}\BibitemShut {Stop}%
\bibitem [{\citenamefont {Kim}\ \emph {et~al.}(2014)\citenamefont {Kim},
  \citenamefont {Ikeda},\ and\ \citenamefont {Huse}}]{Kim2014}%
  \BibitemOpen
  \bibfield  {author} {\bibinfo {author} {\bibfnamefont {H.}~\bibnamefont
  {Kim}}, \bibinfo {author} {\bibfnamefont {T.~N.}\ \bibnamefont {Ikeda}},\
  and\ \bibinfo {author} {\bibfnamefont {D.~A.}\ \bibnamefont {Huse}},\
  }\bibfield  {title} {\bibinfo {title} {{Testing whether all eigenstates obey
  the eigenstate thermalization hypothesis}},\ }\href
  {https://doi.org/10.1103/PhysRevE.90.052105} {\bibfield  {journal} {\bibinfo
  {journal} {Phys. Rev. E}\ }\textbf {\bibinfo {volume} {90}},\ \bibinfo {pages} {052105}
  (\bibinfo {year} {2014})}\BibitemShut {NoStop}%
\bibitem [{Note7()}]{Note7}%
  \BibitemOpen
  \bibinfo {note} {Here, we take $J_y$ negative in (a) because we found that
  the $o(1/\protect \sqrt {N})$ term of Eq.~(\ref {eq:weakETHlike}) is smaller
  for larger $J_x-J_y$.}\BibitemShut {Stop}%
\bibitem [{\citenamefont {Shiraishi}(2019)}]{Shiraishi2019}%
  \BibitemOpen
  \bibfield  {author} {\bibinfo {author} {\bibfnamefont {N.}~\bibnamefont
  {Shiraishi}},\ }\bibfield  {title} {\bibinfo {title} {Proof of the absence of
  local conserved quantities in the {XYZ} chain with a magnetic field},\ }\href
  {https://doi.org/10.1209/0295-5075/128/17002} {\bibfield  {journal} {\bibinfo
   {journal} {Europhys. Lett.}\ }\textbf {\bibinfo {volume}
  {128}},\ \bibinfo {pages} {17002} (\bibinfo {year} {2019})}\BibitemShut
  {NoStop}%
\bibitem [{\citenamefont {Alba}(2015)}]{Alba2015}%
  \BibitemOpen
  \bibfield  {author} {\bibinfo {author} {\bibfnamefont {V.}~\bibnamefont
  {Alba}},\ }\bibfield  {title} {\bibinfo {title} {{Eigenstate thermalization
  hypothesis and integrability in quantum spin chains}},\ }\href
  {https://doi.org/10.1103/PhysRevB.91.155123} {\bibfield  {journal} {\bibinfo
  {journal} {Phys. Rev. B}\
  }\textbf {\bibinfo {volume} {91}},\ \bibinfo {pages} {155123} (\bibinfo
  {year} {2015})}\BibitemShut {NoStop}%
\bibitem [{\citenamefont {Kubo}\ \emph {et~al.}(1991)\citenamefont {Kubo},
  \citenamefont {Toda},\ and\ \citenamefont {Hashitsume}}]{KTH}%
  \BibitemOpen
  \bibfield  {author} {\bibinfo {author} {\bibfnamefont {R.}~\bibnamefont
  {Kubo}}, \bibinfo {author} {\bibfnamefont {M.}~\bibnamefont {Toda}},\ and\
  \bibinfo {author} {\bibfnamefont {N.}~\bibnamefont {Hashitsume}},\
  }\href@noop {} {\emph {\bibinfo {title} {{Statistical Physics II}}}},\
  \bibinfo {edition} {2nd}\ ed.\ (\bibinfo  {publisher} {Springer, Berlin},\
  \bibinfo {year} {1991})\BibitemShut {NoStop}%
\bibitem [{\citenamefont {Hyuga}\ \emph {et~al.}(2014)\citenamefont {Hyuga},
  \citenamefont {Sugiura}, \citenamefont {Sakai},\ and\ \citenamefont
  {Shimizu}}]{Hyuga2014}%
  \BibitemOpen
  \bibfield  {author} {\bibinfo {author} {\bibfnamefont {M.}~\bibnamefont
  {Hyuga}}, \bibinfo {author} {\bibfnamefont {S.}~\bibnamefont {Sugiura}},
  \bibinfo {author} {\bibfnamefont {K.}~\bibnamefont {Sakai}},\ and\ \bibinfo
  {author} {\bibfnamefont {A.}~\bibnamefont {Shimizu}},\ }\bibfield  {title}
  {\bibinfo {title} {{Thermal pure quantum states of many-particle systems}},\
  }\href {https://doi.org/10.1103/PhysRevB.90.121110} {\bibfield  {journal}
  {\bibinfo  {journal} {Phys. Rev. B}\ }\textbf {\bibinfo {volume} {90}},\ \bibinfo {pages} {121110}
  (\bibinfo {year} {2014})}\BibitemShut {NoStop}%
\bibitem [{\citenamefont {Iyer}\ \emph {et~al.}(2015)\citenamefont {Iyer},
  \citenamefont {Srednicki},\ and\ \citenamefont {Rigol}}]{Iyer2015}%
  \BibitemOpen
  \bibfield  {author} {\bibinfo {author} {\bibfnamefont {D.}~\bibnamefont
  {Iyer}}, \bibinfo {author} {\bibfnamefont {M.}~\bibnamefont {Srednicki}},\
  and\ \bibinfo {author} {\bibfnamefont {M.}~\bibnamefont {Rigol}},\ }\bibfield
   {title} {\bibinfo {title} {{Optimization of finite-size errors in
  finite-temperature calculations of unordered phases}},\ }\href
  {https://doi.org/10.1103/PhysRevE.91.062142} {\bibfield  {journal} {\bibinfo
  {journal} {Phys. Rev. E}\ }\textbf {\bibinfo {volume} {91}},\ \bibinfo {pages} {062142}
  (\bibinfo {year} {2015})}\BibitemShut {NoStop}%
\bibitem [{\citenamefont {Kubo}(1957)}]{Kubo1957}%
  \BibitemOpen
  \bibfield  {author} {\bibinfo {author} {\bibfnamefont {R.}~\bibnamefont
  {Kubo}},\ }\bibfield  {title} {\bibinfo {title} {{Statistical-Mechanical
  Theory of Irreversible Processes. I. General Theory and Simple Applications
  to Magnetic and Conduction Problems}},\ }\href
  {https://doi.org/10.1143/JPSJ.12.570} {\bibfield  {journal} {\bibinfo
  {journal} {J. Phys. Soc. Jpn.}\ }\textbf {\bibinfo
  {volume} {12}},\ \bibinfo {pages} {570} (\bibinfo {year} {1957})}\BibitemShut {NoStop}%
\bibitem [{\citenamefont {Pines}\ and\ \citenamefont
  {Nozieres}(1966)}]{Pines1966}%
  \BibitemOpen
  \bibfield  {author} {\bibinfo {author} {\bibfnamefont {D.}~\bibnamefont
  {Pines}}\ and\ \bibinfo {author} {\bibfnamefont {P.}~\bibnamefont
  {Nozieres}},\ }\href@noop {} {\emph {\bibinfo {title} {{The Theory of Quantum
  Liquids, Vol I: Normal Fermi Liquids}}}}\ (\bibinfo  {publisher} {W. A.
  Benjamin, New York},\ \bibinfo {year} {1966})\BibitemShut {NoStop}%
\bibitem [{\citenamefont {Giuliani}\ and\ \citenamefont
  {Vignale}(2005)}]{Giuliani2005}%
  \BibitemOpen
  \bibfield  {author} {\bibinfo {author} {\bibfnamefont {G.}~\bibnamefont
  {Giuliani}}\ and\ \bibinfo {author} {\bibfnamefont {G.}~\bibnamefont
  {Vignale}},\ }\href@noop {} {\emph {\bibinfo {title} {{Quantum Theory of the
  Electron Liquid}}}}\ (\bibinfo  {publisher} {Cambridge University Press,
  Cambridge, England},\ \bibinfo {year} {2005})\BibitemShut {NoStop}%
\bibitem [{\citenamefont {D.N.Zubarev}(1974)}]{Zubarev1974}%
  \BibitemOpen
  \bibfield  {author} {\bibinfo {author} {\bibnamefont {D.N.Zubarev}},\
  }\href@noop {} {\emph {\bibinfo {title} {{Nonequilibrium Statistical
  Thermodynamics}}}}\ (\bibinfo  {publisher} {Consultants Bureau, New York},\
  \bibinfo {year} {1974})\BibitemShut {NoStop}%
\bibitem [{\citenamefont {Zubarev}\ \emph {et~al.}(1996)\citenamefont
  {Zubarev}, \citenamefont {Morozov},\ and\ \citenamefont
  {R\"{o}pke}}]{Zubarev1996}%
  \BibitemOpen
  \bibfield  {author} {\bibinfo {author} {\bibfnamefont {D.}~\bibnamefont
  {Zubarev}}, \bibinfo {author} {\bibfnamefont {V.}~\bibnamefont {Morozov}},\
  and\ \bibinfo {author} {\bibfnamefont {G.}~\bibnamefont {R\"{o}pke}},\
  }\href@noop {} {\emph {\bibinfo {title} {{Statistical Mechanics of
  Nonequilibrium Processes, Volume 1: Basic Concepts, Kinetic Theory}}}} (\bibinfo  {publisher} {Akademie-Verlag,
  Berlin},\ \bibinfo {year} {1996})\BibitemShut {NoStop}%
\bibitem [{\citenamefont {Zubarev}\ \emph {et~al.}(1997)\citenamefont
  {Zubarev}, \citenamefont {Morozov},\ and\ \citenamefont
  {R\"{o}pke}}]{Zubarev1997}%
  \BibitemOpen
  \bibfield  {author} {\bibinfo {author} {\bibfnamefont {D.}~\bibnamefont
  {Zubarev}}, \bibinfo {author} {\bibfnamefont {V.}~\bibnamefont {Morozov}},\
  and\ \bibinfo {author} {\bibfnamefont {G.}~\bibnamefont {R\"{o}pke}},\
  }\href@noop {} {\emph {\bibinfo {title} {{Statistical Mechanics of
  Nonequilibrium Processes, Volume 2: Relaxation and Hydrodynamic
  Processes}}}} (\bibinfo  {publisher}
  {Akademie-Verlag, Berlin},\ \bibinfo {year} {1997})\BibitemShut {NoStop}%
\bibitem [{\citenamefont {Falk}(1968)}]{Falk1968}%
  \BibitemOpen
  \bibfield  {author} {\bibinfo {author} {\bibfnamefont {H.}~\bibnamefont
  {Falk}},\ }\bibfield  {title} {\bibinfo {title} {{Lower Bound for the
  Isothermal Magnetic Susceptibility}},\ }\href
  {https://doi.org/10.1103/PhysRev.165.602} {\bibfield  {journal} {\bibinfo
  {journal} {Phys. Rev.}\ }\textbf {\bibinfo {volume} {165}},\ \bibinfo
  {pages} {602} (\bibinfo {year} {1968})}\BibitemShut {NoStop}%
\bibitem [{\citenamefont {Wilcox}(1968)}]{Wilcox1968}%
  \BibitemOpen
  \bibfield  {author} {\bibinfo {author} {\bibfnamefont {R.~M.}\ \bibnamefont
  {Wilcox}},\ }\bibfield  {title} {\bibinfo {title} {{Bounds for the
  Isothermal, Adiabatic, and Isolated Static Susceptibility Tensors}},\ }\href
  {https://doi.org/10.1103/PhysRev.174.624} {\bibfield  {journal} {\bibinfo
  {journal} {Phys. Rev.}\ }\textbf {\bibinfo {volume} {174}},\ \bibinfo
  {pages} {624} (\bibinfo {year} {1968})}\BibitemShut {NoStop}%
\bibitem [{\citenamefont {Suzuki}(1971)}]{Suzuki1971}%
  \BibitemOpen
  \bibfield  {author} {\bibinfo {author} {\bibfnamefont {M.}~\bibnamefont
  {Suzuki}},\ }\bibfield  {title} {\bibinfo {title} {{Ergodicity, constants of
  motion, and bounds for susceptibilities}},\ }\href
  {https://doi.org/10.1016/0031-8914(71)90226-6} {\bibfield  {journal}
  {\bibinfo  {journal} {Physica}\ }\textbf {\bibinfo {volume} {51}},\ \bibinfo
  {pages} {277} (\bibinfo {year} {1971})}\BibitemShut {NoStop}%
\bibitem [{Note8()}]{Note8}%
  \BibitemOpen
  \bibinfo {note} {For the former order of limits ($\varepsilon \to +0$ after
  $N\to \infty $), the conventional wisdom \cite {Pines1966,Giuliani2005} is
  that $\displaystyle \protect \qopname \relax m{lim}_{\vb *{k} \to \vb *{0}}
  \protect \qopname \relax m{lim}_{\varepsilon \to +0} \protect \qopname \relax
  m{lim}_{N \to \infty }\chi _N^{\protect \text {Kubo}}(\vb *{k},
  0+i\varepsilon ) = \chi _\infty ^{T}(\vb *{0})$, which however does not
  always hold. Our results (ii)-(iv) suggest the condition for the validity of
  this wisdom, although we have not yet proved $\displaystyle \protect \qopname
  \relax m{lim}_{\varepsilon \to +0} \protect \qopname \relax m{lim}_{N \to
  \infty }\chi _N^{\protect \text {Kubo}}(\vb *{k}, 0+i\varepsilon ) = \chi
  _\infty ^{\protect \text {qch}}(\vb *{k})$.}\BibitemShut {Stop}%
\end{thebibliography}
%

\clearpage

\pagestyle{empty}
\parindent=0mm
\setlength{\textwidth}{1.15\textwidth}

\begin{figure*}
\vspace{-10mm}
\hspace{-28mm}
\includegraphics[page=1,clip]{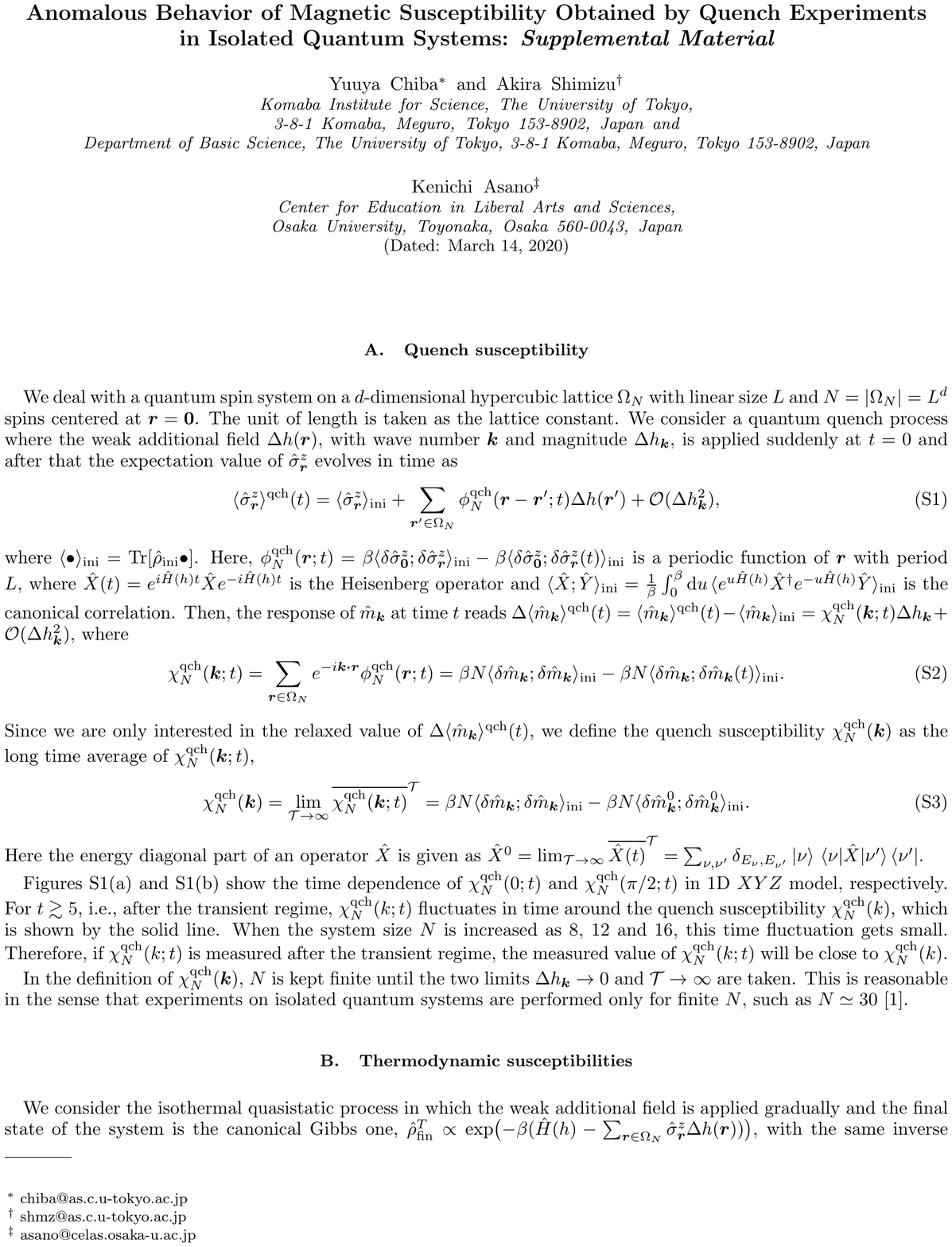}
\end{figure*}

\begin{figure*}
\vspace{-10mm}\hspace{-28mm}
\includegraphics[page=2,clip]{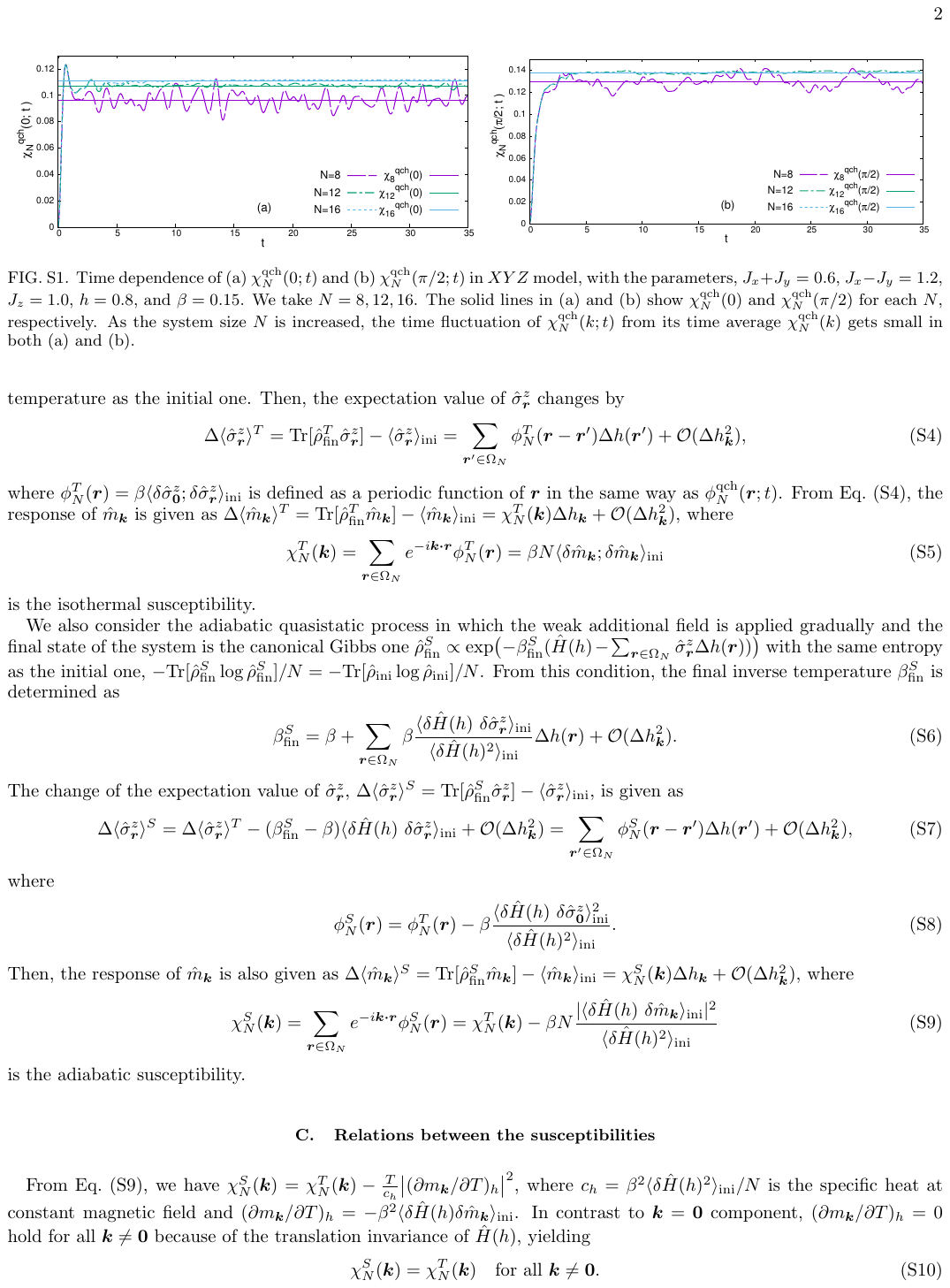}
\end{figure*}

\begin{figure*}
\vspace{-10mm}\hspace{-28mm}
\includegraphics[page=3,clip]{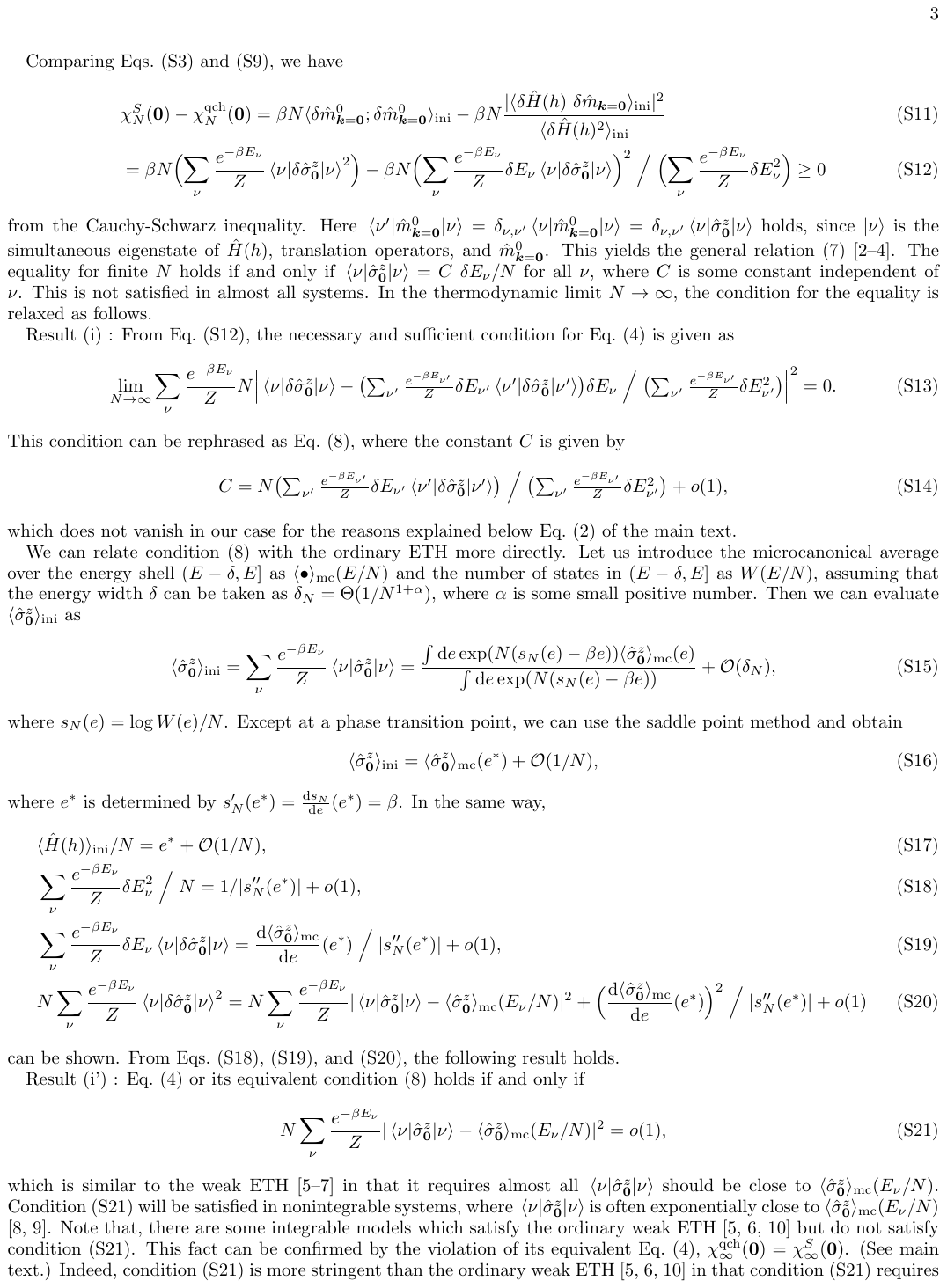}
\end{figure*}

\begin{figure*}
\vspace{-10mm}
\hspace{-28mm}
\includegraphics[page=4,clip]{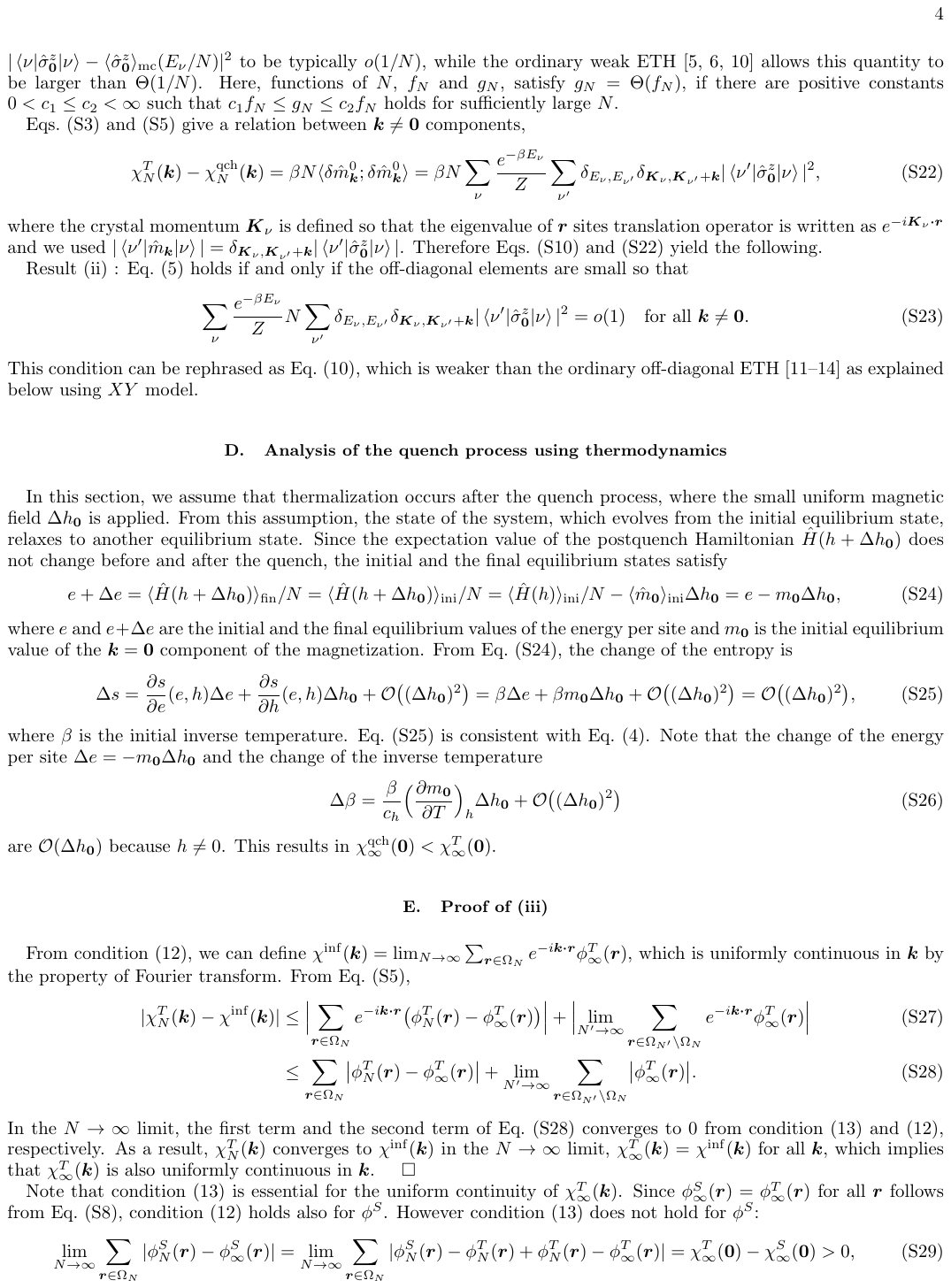}
\end{figure*}

\begin{figure*}
\vspace{-10mm}\hspace{-28mm}
\includegraphics[page=5,clip]{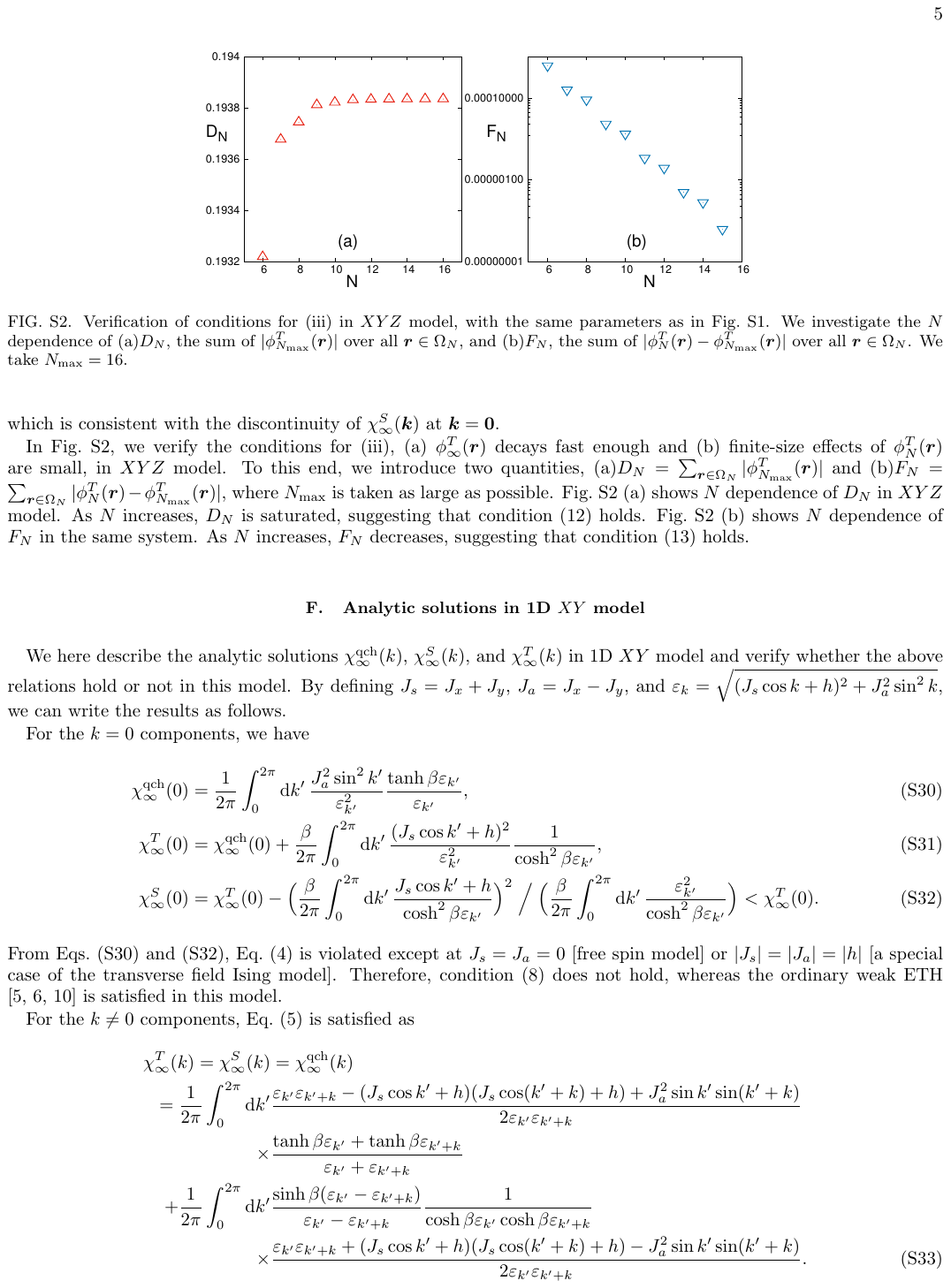}
\end{figure*}

\begin{figure*}
\vspace{-10mm}\hspace{-28mm}
\includegraphics[page=6,clip]{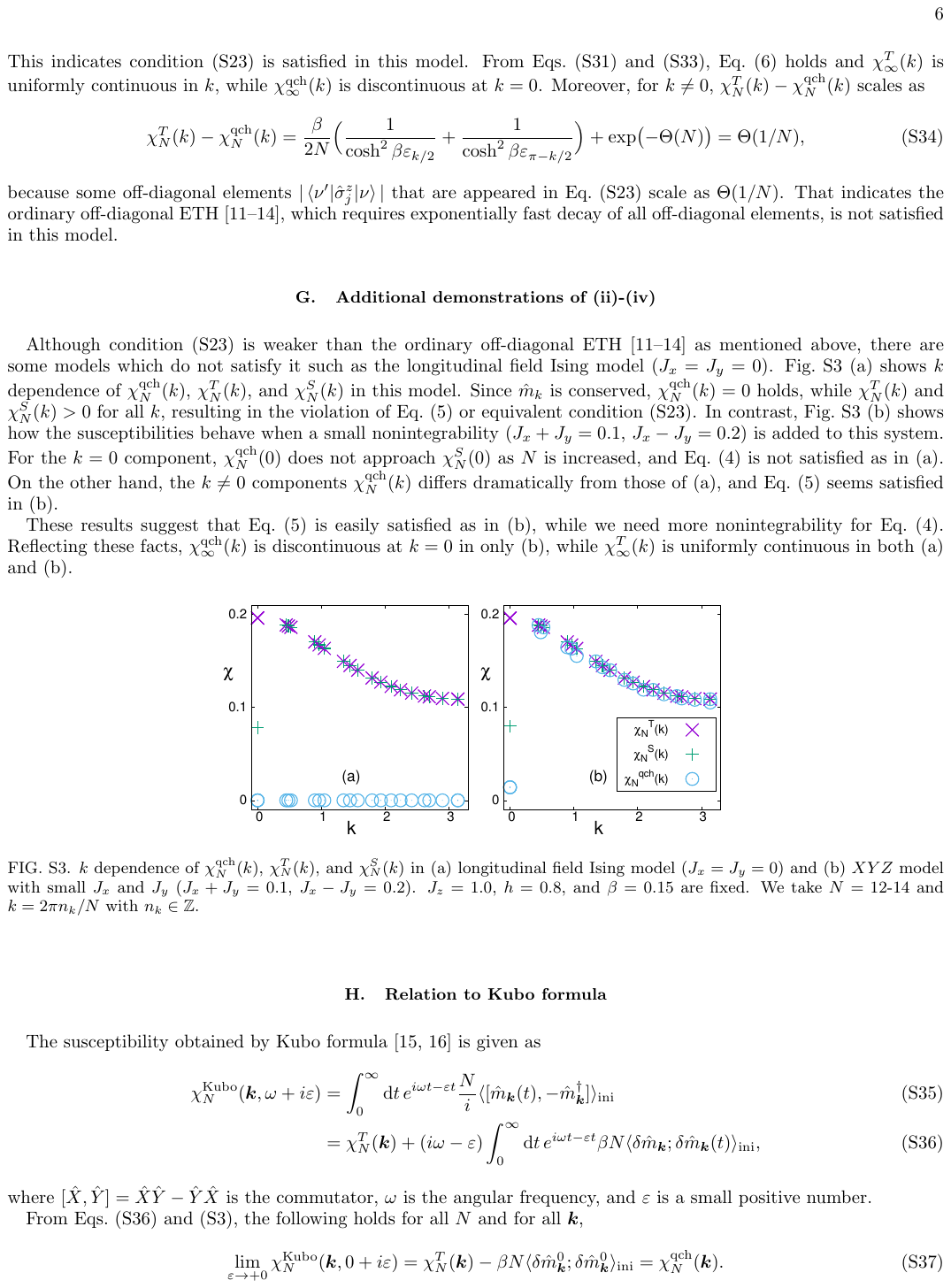}
\end{figure*}

\begin{figure*}
\vspace{-10mm}\hspace{-28mm}
\includegraphics[page=7,clip]{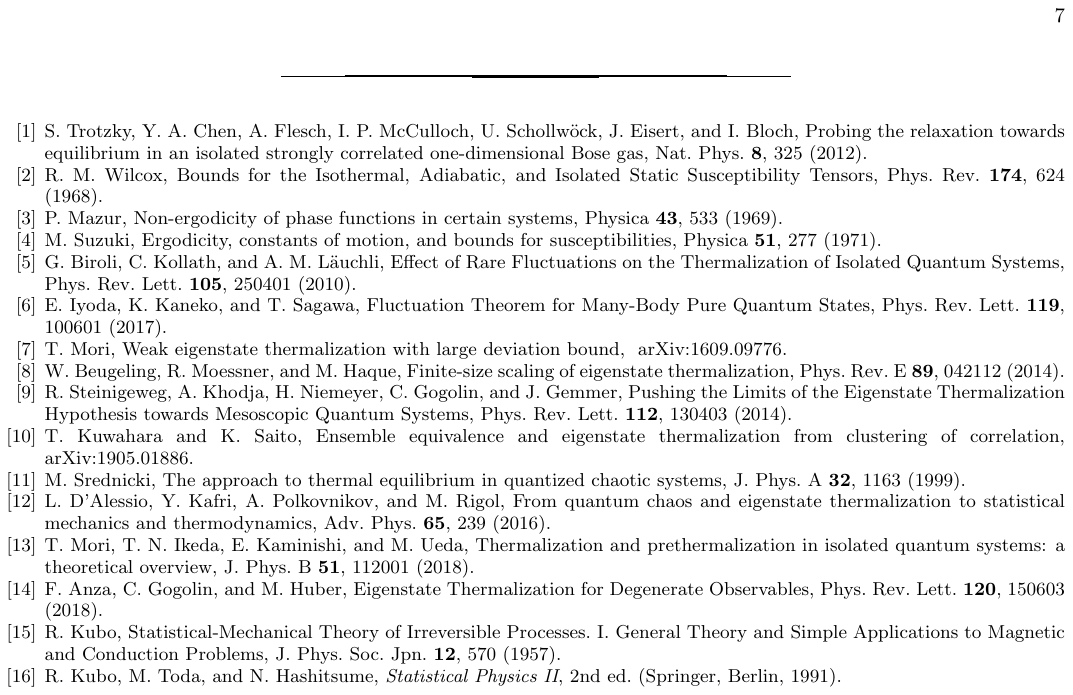}
\end{figure*}

\end{document}